\documentclass[preprint,aps,preprintnumbers,amsmath,amssymb,nofootinbib,epsf,epsfig]
{revtex4}
\usepackage[utf8]{inputenc}
\usepackage{graphicx}
\usepackage{amsmath}
\usepackage{amssymb}
\usepackage{array}
\usepackage{fancyhdr}
\usepackage{float}
\usepackage{booktabs}
\usepackage{multirow}
\usepackage{color}
\usepackage{caption}
\usepackage{subcaption}
\usepackage{natbib}
\usepackage{pifont}
\usepackage[colorlinks, citecolor=cyan]{hyperref}
\usepackage{epsfig}
\usepackage{slashed}
\usepackage{appendix}
\usepackage{comment}

\newcommand{\abs}[1]{\left| #1 \right|}
\begin{document}

\newcommand{\be}{\begin{equation}}
\newcommand{\ee}{\end{equation}}
\newcommand{\bea}{\begin{eqnarray}}
\newcommand{\eea}{\end{eqnarray}}
\newcommand{\nn}{\nonumber}
\def\ds{\displaystyle}
\def\s1{\hat s}
\def\para{\parallel}
\newcommand{\mrm}[1]{\mathrm{#1}}
\newcommand{\mc}[1]{\mathcal{#1}}
\def\CP{{\it CP}~}
\def\cp{{\it CP}}
\def\ml{m_\mu}
\title{\large Revealing the $q^2$ dependence in $b \to s \mu^+ \mu^-$ baryonic  decay modes}

\author{Ajay Kumar Yadav$^a$}
\email{yadavajaykumar286@gmail.com}
\author{Manas Kumar Mohapatra$^b$} 
\email{manasmohapatra12@gmail.com} 
\author{Suchismita Sahoo$^a$}
\email{suchismita8792@gmail.com}
\affiliation{Department of Physics, Central University of Karnataka, Kalaburagi-585367, India $^a$ \\ School of Physics, University of Hyderabad, Hyderabad-500046, India $^b$}
\begin{abstract}
Measurements from the LHCb experiment and $B$ factories have revealed several discrepancies in angular observables of rare semileptonic $B$ decays involving the quark level transition $b \to s \ell^+ \ell^-$. In this work, we conduct a model independent comparative analysis of the rare semileptonic decays of  baryons $\Lambda_b$, $\Sigma_b$ and $\Xi_b$, exploring various new physics scenarios. Our analysis includes predictions for branching ratios and angular observables, including forward backward asymmetry, longitudinal polarization fractions, and lepton flavor universality ratios, both within the Standard Model and across six distinct new physics scenarios. Notably, we find that certain new physics scenarios significantly affect the  observables in the $(\Lambda_b, \Sigma_b, \Xi_b) \to (\Lambda, \Sigma, \Xi) \mu^+ \mu^-$ processes.
\end{abstract}
\maketitle

\section{Introduction}

The rare process induced by the flavor changing neutral currents (FCNC) in the $b \to s \ell \ell$ transition serves as a critical testing ground for the Standard Model (SM) at loop level and is highly sensitive to new physics (NP) effects. Recent investigations into the $b \to s \mu^+ \mu^-$ transitions have been particularly compelling due to anomalies reported by the LHCb and Belle II Collaborations~\cite{LHCb:2017avl,Belle:2016fev,CMS:2024syx}. Noteworthy among these anomalies are the ratios $\mathcal{R}_K$ and $\mathcal{R}_{K^*}$~\cite{Hiller:2003js}, defined as  
\begin{equation}  
\mathcal{R}_{K^{(*)}} = \frac{\text{BR}(B \to K^{(*)} \mu \mu)}{\text{BR}(B \to K^{(*)} e e)}\,,  
\end{equation}  
which are anticipated to be unity in the SM, embodying the principle of lepton flavor universality (LFU). These ratios are especially valuable due to their theoretical cleanliness, as hadronic uncertainties are expected to cancel out. Recent measurements from LHCb~\cite{LHCb:2022qnv, LHCb:2022vje} are consistent with the SM predictions for these ratios. However, other observables within $b \to s \ell \ell$ transitions, such as $P_5'$ and various branching fractions, display significant deviations from SM expectations. For example, deviations of $3.3\sigma$ in $P_5'$ have been reported by LHCb~\cite{LHCb:2013ghj, LHCb:2015svh} and ATLAS~\cite{ATLAS:2018gqc}, while the branching ratio for $B_s \to \phi \mu^+ \mu^-$ exhibits a $3.3\sigma$ deviation~\cite{LHCb:2021zwz, LHCb:2021xxq, LHCb:2015wdu} in the $q^2$ range of $[1.1, 6.0]$ GeV$^2$. Deviations of $1.4\sigma$ and $1.5\sigma$ have also been observed in the measurements of $\mathcal{R}_{K_S^0}$ and $\mathcal{R}_{K^{*+}}$~\cite{LHCb:2021lvy}, respectively. Moreover, other observables such as forward backward asymmetries, longitudinal polarization asymmetries, CP averaged observables, and CP asymmetries in $B \to K^{(*)}\mu^+\mu^-$ decays across various $q^2$ intervals show deviations from SM predictions~\cite{LHCb:2020dof,LHCb:2020lmf, LHCb:2019vks, LHCb:2014cxe,CMS-PAS-BPH-21-002,BaBar:2012mrf, LHCb:2020gog, Mahmoudi:2024zna}.  These discrepancies could potentially be due to new physics beyond the SM (BSM), a lack of comprehensive understanding of hadronic uncertainties, or limitations in the analysis of experimental data. Although the potential for BSM effects is intriguing, it is crucial to first enhance the theoretical framework addressing the substantial hadronic contributions in rare $b \rightarrow s$ transitions.

To complement the mesonic analyses and probe the same quark level transition, baryonic decay channels such as $\Lambda_b \to \Lambda \mu^+ \mu^-$ have emerged as powerful alternatives, offering distinct advantages. While experimental efforts have traditionally focused on mesonic modes, semileptonic baryon decays of the form $B_1 \rightarrow B_2 \ell^+ \ell^-$ present a theoretically rich and experimentally viable avenue to further investigate the observed anomalies. Moreover, approximately 20\% of the total number of hadrons produced at LHCb~\cite{LHCb:2011leg, LHCb:2014ofc} are $\Lambda_b$ baryons. Several compelling reasons make the study of baryonic decays, particularly those involving the $\Lambda_b$ baryon, highly appealing compared to their mesonic counterparts:

\begin{itemize}
    \item The spin 1/2 nature of the initial and final state baryons allows for a richer angular analysis. Just like mesonic $B\to K^{*} l^+ l^-$ case, the unpolarized $\Lambda_b \to \Lambda(\to p\pi^-)\ell^+\ell^-$ features 10 angular observables, which increase to 34 in the case of polarized $\Lambda_b$ production, substantially more than the 8--10 observables available in $B \to K^*\ell^+\ell^-$ decays.
    
    \item The $\Lambda$ baryon undergoes a weak, parity violating decay to $p\pi^-$, enabling direct access to hadron-side forward backward asymmetries and CP odd observables, which remain inaccessible in mesonic decays that end in strong final states.

    \item Unlike $B \to K^*$ decays, where the $K^*$ is an unstable vector meson requiring complicated treatments of multi-hadron states, the $\Lambda$ is stable under strong interactions. This allows the form factors for $\Lambda_b \to \Lambda$ to be calculated more reliably using lattice QCD without resorting to complex finite volume techniques. This reduces theoretical uncertainties compared to $B \to K^*$ transitions, which suffer from light cone sum rule limitations and significant nonlocal effects.

    \item Baryonic decays may be less sensitive to nonlocal charm loop contributions and long-distance effects that plague mesonic observables such as $P_5'$.

    \item These decays offer enhanced sensitivity to certain Wilson coefficients, particularly those associated with right-handed currents and tensor operators due to the distinct helicity structure of baryons.

    \item The $\Lambda_b$ is relatively long-lived and abundantly produced at LHCb, making such decay channels experimentally accessible. Indeed, $\Lambda_b \to \Lambda \mu^+ \mu^-$ has already been observed and studied by CDF and LHCb~\cite{CDF:2011buy, LHCb:2013uqx, LHCb:2015tgy}.

    \item In scenarios involving violation of lepton flavor universality, baryonic channels like $\Lambda_b \to \Lambda \ell^+ \ell^-$ (with $\ell = e, \mu, \tau$) serve as theoretically clean and experimentally promising probes of new physics, complementary to their mesonic counterparts.

    \item The complex angular structure allows for the construction of optimized observables that are less sensitive to hadronic form factors, thereby enhancing the robustness of NP interpretations.
\end{itemize}

In this light, baryonic decay modes involving the $b \to s \ell^+ \ell^-$ quark-level transition have garnered significant attention, prompting extensive theoretical investigation into processes such as $\Lambda_b \to \Lambda \mu^+ \mu^-$, $\Sigma_b \to \Sigma \mu^+ \mu^-$, and $\Xi_b \to \Xi \mu^+ \mu^-$. The semileptonic decay $\Lambda_b^0 \to \Lambda \mu^+ \mu^-$ was first observed by the CDF Collaboration at Fermilab using data collected with the CDF II detector at a center-of-mass energy of $\sqrt{s} = 1.96$ TeV~\cite{CDF:2011buy}. The observation had a statistical significance of $5.8\sigma$ based on 24 signal events, with a measured branching fraction of $\mathcal{B}(\Lambda_b^0 \to \Lambda \mu^+ \mu^-) = [1.73 \pm 0.42\,(\text{stat}) \pm 0.55\,(\text{syst})] \times 10^{-6}$. Subsequently, the LHCb Collaboration also observed this rare decay using data corresponding to an integrated luminosity of $1.0~\mathrm{fb}^{-1}$ at $\sqrt{s} = 7$ TeV~\cite{LHCb:2013uqx}, reporting a signal yield of $78 \pm 12$ events and a branching fraction of $\mathcal{B} = [0.96 \pm 0.16\,(\text{stat}) \pm 0.13\,(\text{syst}) \pm 0.21\,(\text{norm})] \times 10^{-6}$. An updated measurement of the differential branching fraction has been reported by LHCb using $pp$ collision data corresponding to an integrated luminosity of $3.0~\mathrm{fb}^{-1}$ collected at $\sqrt{s} = 7$ and $8$ TeV~\cite{LHCb:2015tgy}. The analysis uses non-overlapping and theoretically motivated $q^2$ intervals, normalized to the tree-level decay $\Lambda_b^0 \to J/\psi\,\Lambda$. In addition to differential branching fractions, LHCb also presented the first angular analysis of this decay, including a measurement of the lepton forward–backward asymmetry.  Recent searches by the LHCb collaboration, such as $\Sigma^+ \to p \mu^+ \mu^-$ using $3~\text{fb}^{-1}$ of data at $\sqrt{s} = 7$ and $8$~TeV~\cite{LHCb:2017rdd}, and $\Xi_b^- \to \Xi^- \gamma$ using $5.4~\text{fb}^{-1}$ at $\sqrt{s} = 13$~TeV~\cite{LHCb:2021hfz}, demonstrate the experimental accessibility of rare baryonic decays. These developments indicate that the modes $\Sigma_b \to \Sigma \mu^+ \mu^-$ and $\Xi_b \to \Xi \mu^+ \mu^-$ may also become experimentally accessible at LHCb in the near future. In this context, theoretical predictions are crucial for guiding and motivating experimental efforts. A wide range of studies has explored these decay channels in detail~\cite{Katirci:2012eh, Azizi:2011mw, Aliev:2010uy, Azizi:2013eta, Gutsche:2013pp, Boer:2014kda, Detmold:2016pkz, Faustov:2016pal, Das:2018iap, Das:2022xjg, Sahoo:2016nvx, Mohanta:2010eb, Kumar:2015tnz, Tan:2023opd, Wang:2021uzi, Erben:2022tdu}, offering valuable insights into the underlying dynamics and new physics sensitivity.

\noindent
Against this backdrop, in this study, we investigate the implications of a model-independent effective field theory framework on exclusive $b \to s \mu^+ \mu^-$ baryonic decay modes. Focusing on vector and axial-vector operators, we extract constraints on the corresponding Wilson coefficients by performing global fits to current experimental data. The analysis incorporates observables from both leptonic $B_s \to \mu^+ \mu^-$ and semileptonic $B \to (K^{(*)}, \phi) \mu^+ \mu^-$ decays, including branching ratios, forward-backward asymmetries, longitudinal polarization asymmetries, lepton flavor universality ratios, and form factor-independent quantities~\cite{Mohapatra:2024lmp}. Using the resulting fits, we present updated predictions for the rare baryonic decays $(\Lambda_b, \Sigma_b, \Xi_b) \to (\Lambda, \Sigma, \Xi) \mu^+ \mu^-$ within both the SM and various NP scenarios. These predictions allow for a detailed exploration of the phenomenological consequences of NP in the baryonic sector. By transferring constraints from mesonic to baryonic transitions, we perform critical consistency checks across different hadronic environments, thereby assessing the robustness and universality of possible NP explanations.

This paper is organized as follows: Section II provides an overview of the effective Hamiltonian relevant to semileptonic decays involving $b \to s \ell^+ \ell^-$ quark-level transitions and details the expressions for various observables in the $B_1 \to B_2 \mu^+ \mu^-$ process. Section III focuses on the numerical fitting of new (axial)vector Wilson coefficients within both one-dimensional and two-dimensional scenarios and examines the impact of these newly fitted parameters on $b \to s \mu^+ \mu^-$ mediated baryonic decay observables. Finally, Section IV presents a summary and conclusion of our findings.

\section{Theoretical Framework}
\subsection{Effective Hamiltonian}
The baryonic decay \( B_1 \rightarrow B_2 \ell^+ \ell^- \), where $B_1=\Lambda_b, \Sigma_b, \Xi_b$ and $B_2=\Lambda, \Sigma, \Xi$, arises from a flavor-changing neutral current (FCNC) transition, described by the SM effective Hamiltonian, similar to other \( b \rightarrow s\, \ell^+ \ell^- \) decays, and can be expressed as \cite{Buchalla:1995vs, Ali:1999mm, Altmannshofer:2008dz} 
\begin{align*}
    \mathcal{H}_{eff}=\frac{-4 G_F}{\sqrt{2}}\,V_{tb}\,V_{ts}^*\,\Big[\sum_{i=1}^{6}\,C_i(\mu)\,\mathcal{O}_i+C_7^{(\prime)}\frac{e}{16\,\pi^2}\,(\overline{s}\sigma_{\mu\nu}(m_s\,P_L+m_b\,P_R)b)\,F^{\mu\nu}\\
    +C_9^{(\prime)}\frac{\alpha_e}{4\pi}(\overline{s}\gamma^{\mu}P_L\,b)\Bar{\ell}\gamma_{\mu}\,\ell+C_{10}^{(\prime)}\frac{\alpha_e}{4\pi}(\overline{s}\gamma^{\mu}P_L\,b)\Bar{\ell}\gamma_{\mu}\gamma_5\,\ell\Big]\,.
\end{align*}
Here, \( G_F \) (\( \alpha_e \)) represents the Fermi coupling (fine structure constant), and the CKM matrix elements are denoted by \( V_{tb}\, V_{ts}^* \).
The relevant four-fermion operators, $\mathcal{O}_i^{(\prime)}$ for $i = 7, 9, 10$, are explicitly defined as
\begin{eqnarray*}
     \mathcal{O}_7 &=& \frac{e}{16 \pi^2} m_b (\bar{s} \sigma_{\mu \nu} P_R b) F^{\mu \nu}, \quad
     \mathcal{O}_7' = \frac{e}{16 \pi^2} m_b (\bar{s} \sigma_{\mu \nu} P_L b) F^{\mu \nu}, \\
     \mathcal{O}_9 &=& \frac{e^2}{16 \pi^2} (\bar{s} \gamma_{\mu} P_L b) (\bar{\mu} \gamma^{\mu} \mu), \quad
     \mathcal{O}_9' = \frac{e^2}{16 \pi^2} (\bar{s} \gamma_{\mu} P_R b) (\bar{\mu} \gamma^{\mu} \mu), \\
     \mathcal{O}_{10} &=& \frac{e^2}{16 \pi^2} (\bar{s} \gamma_{\mu} P_L b) (\bar{\mu} \gamma^{\mu} \gamma_5 \mu), \quad
     \mathcal{O}_{10}' = \frac{e^2}{16 \pi^2} (\bar{s} \gamma_{\mu} P_R b) (\bar{\mu} \gamma^{\mu} \gamma_5 \mu),
\end{eqnarray*}

where $P_{R,L} = \frac{1 \pm \gamma_5}{2}$ are the chirality projection operators, and the $C_i^{(\prime)}$'s are the corresponding Wilson coefficients \cite{Ali:1999mm}. 
The couplings \( C_{i=1,\dots,6} \) correspond to the Wilson coefficients associated with current-current operators \((i=1,2)\) and QCD penguin operators \((i=3,\dots,6)\). The couplings \( C_7 \), (\( C_9^{(\prime)} \), \( C_{10}^{(\prime)} \)) pertain to radiative (electromagnetic) penguin operators.
It is worth noting that the SM does not contribute to the primed Wilson coefficients. These operators only emerge in extensions of the SM.

\subsection{Observables of $B_1 \to B_2 \ell^+ \ell^-$ decays}

The angular distribution of the baryonic decay \( B_1 \to B_2 \) can be parameterized in terms of \( q^2 \), \( \theta_{B_2} \), \( \theta_l \), and \( \phi \), where \( q^2 \) denotes the square of the invariant mass of the lepton pair, \( \theta_{B_2} \) and \( \theta_l \) are the helicity angles of the hadronic and leptonic subsystems, respectively, and \( \phi \) is the azimuthal angle between the two planes. The four-fold angular distribution for the \( B_1 \to B_2 \ell^+ \ell^- \) process is expressed as \cite{Gutsche:2013pp, Boer:2014kda}
\begin{align}{\label{ang dist}}
    \frac{d^4\Gamma}{dq^2 d\cos{\theta_l} d\cos{\theta_{B_2}}d\phi}=\frac{3}{8\,\pi}\mathcal{K}(q^2,\cos{\theta_l},\cos{\theta_{B_2}},\phi),
    \end{align} 
where $\mathcal{K}(q^2,\cos{\theta_l},\cos{\theta_{B_2}},\phi)$, described in terms of angular coefficients $(\mathcal{K}_i)$, is expressed as
    \begin{align*}
        \mathcal{K}(q^2,\cos{\theta_l},\cos{\theta_{B_2}},\phi)= &\mathcal{K}_{1ss}\,\sin^2{\theta_l}+\mathcal{K}_{1cc}\,\cos^2{\theta_l}+\mathcal{K}_{1c}\,\cos{\theta_l}+\\&(\mathcal{K}_{2ss}\,\sin^2{\theta_l}+\mathcal{K}_{2cc}\,\cos^2{\theta_l}+\mathcal{K}_{2c}\,\cos{\theta_l})\,\cos{\theta_{B_2}}+ \\&(\mathcal{K}_{3sc}\,\sin{\theta_l}\cos{\theta_l}+\mathcal{K}_{3s}\,\sin{\theta_l})\sin{\theta_{B_2}\cos{\phi}}+\\&(\mathcal{K}_{4sc}\,\sin{\theta_l}\cos{\theta_l}+\mathcal{K}_{4s}\,\sin{\theta_l})\sin{\theta_{B_2}\sin{\phi}}.
    \end{align*}
The detailed expressions for the angular coefficients are provided in Appendix \ref{A}.  From the differential decay distributions given in Eq. (\ref{ang dist}), one can construct the following observables:

\ding{43} The differential decay rate with $q^2$ \cite{Gutsche:2013pp, Boer:2014kda}: 
\begin{eqnarray}
 \frac{d\Gamma}{dq^2}=2\mathcal{K}_{1ss}+\mathcal{K}_{1cc}\,.
\end{eqnarray}

\ding{43} The lepton flavor universality violation observable:
\begin{eqnarray}
    \mathcal{R}_{B_2}(q^2) = \frac{d\Gamma/dq^2|_{\mu}}{d\Gamma/dq^2|_{e}}\,.
\end{eqnarray}

\ding{43} The forward-backward asymmetry \cite{Boer:2014kda,Detmold:2016pkz}:
\begin{itemize}
     \item   Leptonic: 
     \begin{align}
     \mathcal{A}^\ell_{FB}=\frac{3}{2}\frac{\mathcal{K}_{1c}}{2\mathcal{K}_{1ss}+\mathcal{K}_{1cc}}\,,    
     \end{align}
     
        \item Leptonic - Hadronic: 
        \begin{align}
        \mathcal{A}^{\ell,B_2}_{FB}=\frac{3}{4}\frac{\mathcal{K}_{2c}}{2\mathcal{K}_{1ss}+\mathcal{K}_{1cc}}\,,  
        \end{align}
    \item Hadronic:
    \begin{align}
    \mathcal{A}^{B_2}_{FB}=\frac{1}{2}\frac{2\mathcal{K}_{2ss}+\mathcal{K}_{2cc}}{2\mathcal{K}_{1ss}+\mathcal{K}_{1cc}}\,.    
    \end{align}
        \end{itemize}
\ding{43} The longitudinal polarization asymmetry \cite{Gutsche:2013pp, Boer:2014kda,Detmold:2016pkz}:
\begin{align}
   \mathcal{F}_L=&\frac{2\mathcal{K}_{1ss}-\mathcal{K}_{1cc}}{2\mathcal{K}_{1ss}+\mathcal{K}_{1cc}}\,.
\end{align}
By definition, the transverse polarization fraction is $\mathcal{F}_T=1-\mathcal{F}_L$. 

\subsection{Inputs}
Prior to commencing our analysis, we outline the key input parameters utilized in the numerical computation. For the masses of the leptons, mesons and baryons, the lifetime parameters, and the CKM matrix elements, we use Ref.~\cite{ParticleDataGroup:2024cfk}. For the baryonic form factors, we utilize Lattice QCD calculations for the $\Lambda_b \to \Lambda$ transition, as detailed in Ref. \cite{Detmold:2016pkz}. For the $\Sigma_b \to \Sigma$ and $\Xi_b \to \Xi$ transitions, we adopt the Light-Cone QCD Sum Rules method, following Refs. \cite{Katirci:2012eh, PhysRevD.99.015015} and \cite{Azizi:2011mw}, respectively. Further details on the form factors are provided in Appendix \ref{B}.

\section{New physics sensitivity}
\subsection{Constraint on new physics coupling(s)}
To investigate the observed discrepancies in $b \to s \mu^+ \mu^-$ transitions, we employ a model-independent framework for a thorough data analysis. We focus on observables in \(b \to s \mu^+ \mu^-\) decays that exhibit deviations from the SM predictions. Particular emphasis is placed on scenarios involving the vector and axial-vector Wilson coefficients \(C_9^{(\prime) \mathrm{NP}}\) and \(C_{10}^{(\prime) \mathrm{NP}}\), which may indicate contributions from new physics. In this context, we utilize the new physics coupling parameters corresponding to the most favorable $p$-values, as determined in our previous work, which incorporates the updated lepton flavor universality ratios along with other $b \to s \mu^+ \mu^-$ observables~\cite{Mohapatra:2024lmp}. Specifically, we examine six scenarios: two one-dimensional and four two-dimensional that yield the best p-values and large pull values. Detailed information is provided in Table \ref{tab:1D-2Dfits}.
\begin{table}[htbp]
\centering
\caption{Best-fit value(s) $[1\sigma]$, pull and p-value($\%$) for 1D and 2D new physics scenarios.} \label{tab:1D-2Dfits}
\begin{tabular}{@{}ccccr@{}}
\toprule[1.2pt] 
Scenario & Coefficient & Best fit value [$1\sigma$] & Pull &  p-value ($\%$)  \\ 
\hline
\hline
 \midrule 
\hline \hline
 \textbf{S - A} & $C_9^{\rm NP}$ & $ -0.46$ $[\substack{-0.33 \\ -0.60}]$    & 2.65 & 59.0\\ 
 \textbf{S - B} & $C_9^{\rm NP}=-C_{10}^{\rm NP}$ &   $ -0.19 $  $[\substack{-0.12 \\ -0.25}]$&   2.92 & 66.0 \\
 \textbf{S - C} & $(C_9^{\rm NP},C_{10}^{\rm NP})$ & $(-0.40 [\substack{-0.21 \\ -0.59}], 0.10[\substack{0.23 \\ -0.03}])$ &  3.45& 54.0 \\ 
\textbf{S - D} &  $(C_9^{\rm NP},C_{10}^{'\rm NP})$ &  ($-0.49[\substack{-0.32 \\ -0.66}]$, $-0.17[\substack{-0.06 \\ -0.27}]$ )  & 3.81 & 61.0\\ 
\textbf{S - E} & $(C_{9}^{\rm NP}=C_{9}^{'\rm NP}, C_{10}^{\rm NP}=-C_{10}^{'\rm NP})$ &  ($-0.22[\substack{-0.15 \\ -0.32}] $, $ 0.15 [\substack{0.21 \\ 0.07}]$)  & 3.18 & 51.0 \\ 
\textbf{S - F} & $(C_{9}^{\rm NP}=-C_{10}^{\rm NP},C_{9}^{'\rm NP}=C_{10}^{'\rm NP})$ & ($-0.17[\substack{-0.10 \\ -0.24}] $, $ -0.09 [\substack{0.05 \\ -0.24}]$)  & 3.33 & 53.0\\ 

\bottomrule[1.6pt] 
\end{tabular}
\end{table}
\subsection{Results and Discussion}
We have carried out a comprehensive theoretical and numerical analysis, yielding several significant insights. The results are presented for six distinct new physics scenarios: $\mathbf{S\text{-}A}$, $\mathbf{S\text{-}B}$, $\mathbf{S\text{-}C}$, $\mathbf{S\text{-}D}$, $\mathbf{S\text{-}E}$, and $\mathbf{S\text{-}F}$. 
\begin{figure}[htb]
    \centering
    \includegraphics[width=0.85\linewidth]{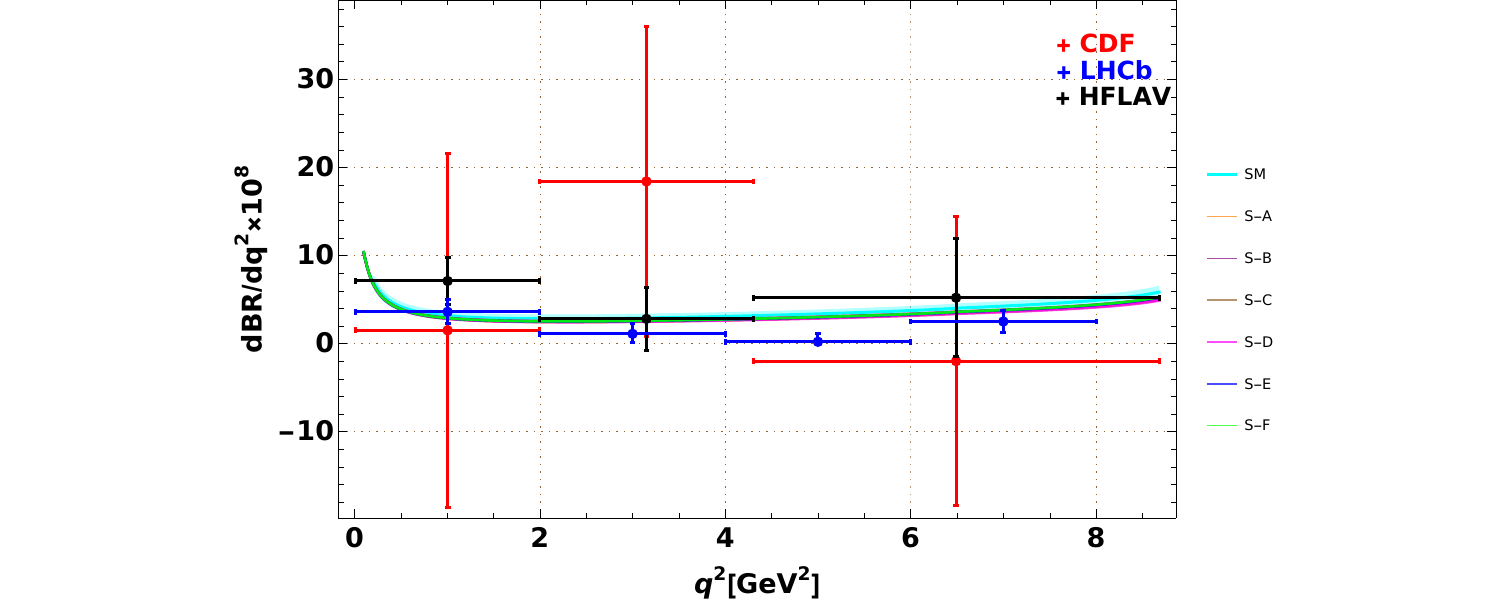}
    \caption{The $q^2$ dependence of the differential branching ratio for the decay $\Lambda_b \to \Lambda \mu^+ \mu^-$ in the SM and various NP scenarios. The available experimental measurements from LHCb, CDF, and the HFLAV averages are shown in blue, red, and black, respectively.}
    \label{Fig:BR-Exp-cont}
\end{figure}
Figure~\ref{Fig:BR-Exp-cont} illustrates the $q^2$ dependence of the differential branching ratio for the decay $\Lambda_b \to \Lambda \mu^+ \mu^-$ in the low-$q^2$ region, $q^2 \in [0,\,8.43]~\mathrm{GeV}^2$, both within the SM and across all the considered NP scenarios. The available experimental measurements from LHCb~\cite{LHCb:2015tgy}, CDF~\cite{CDF:2011buy}, and the HFLAV averages~\cite{TheHeavyFlavorAveragingGroup:2025ung}  are shown in blue, red, and black, respectively.
\begin{figure}[H]
    \centering
  \hspace*{-1.7cm}\includegraphics[width=18cm,height=6cm]{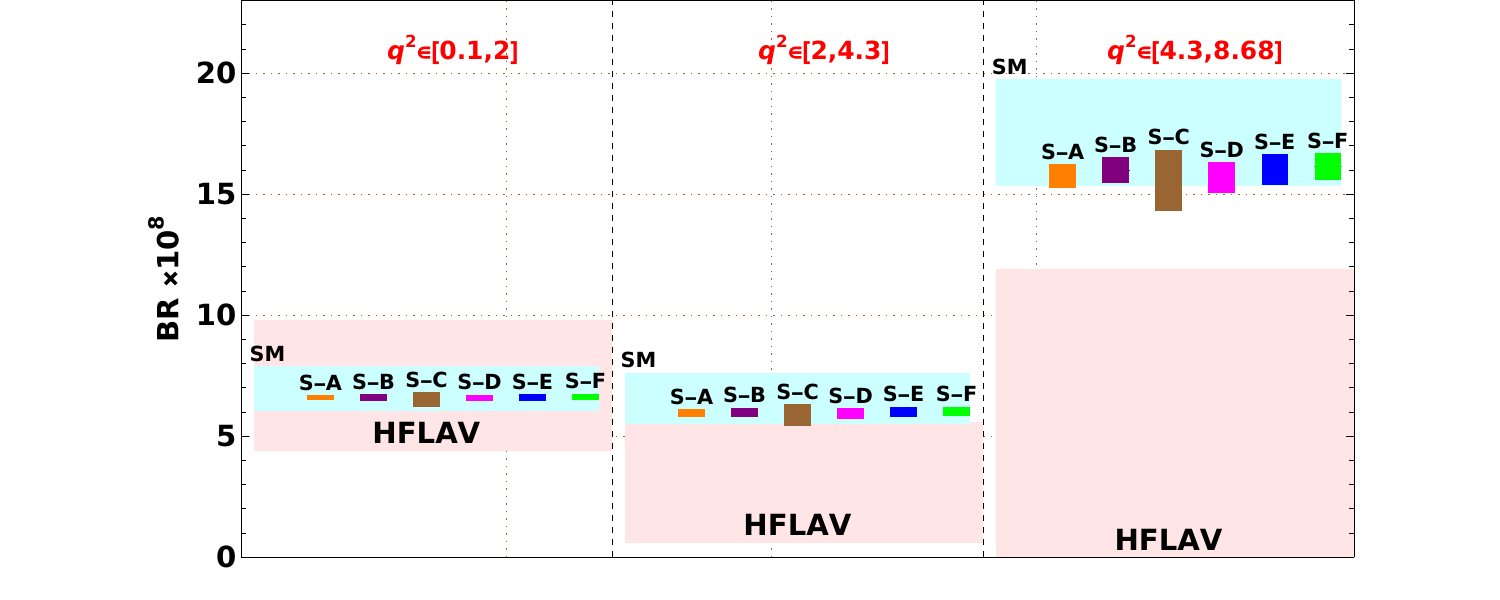}
  \hspace*{-1.7cm}\includegraphics[width=18cm,height=6cm]{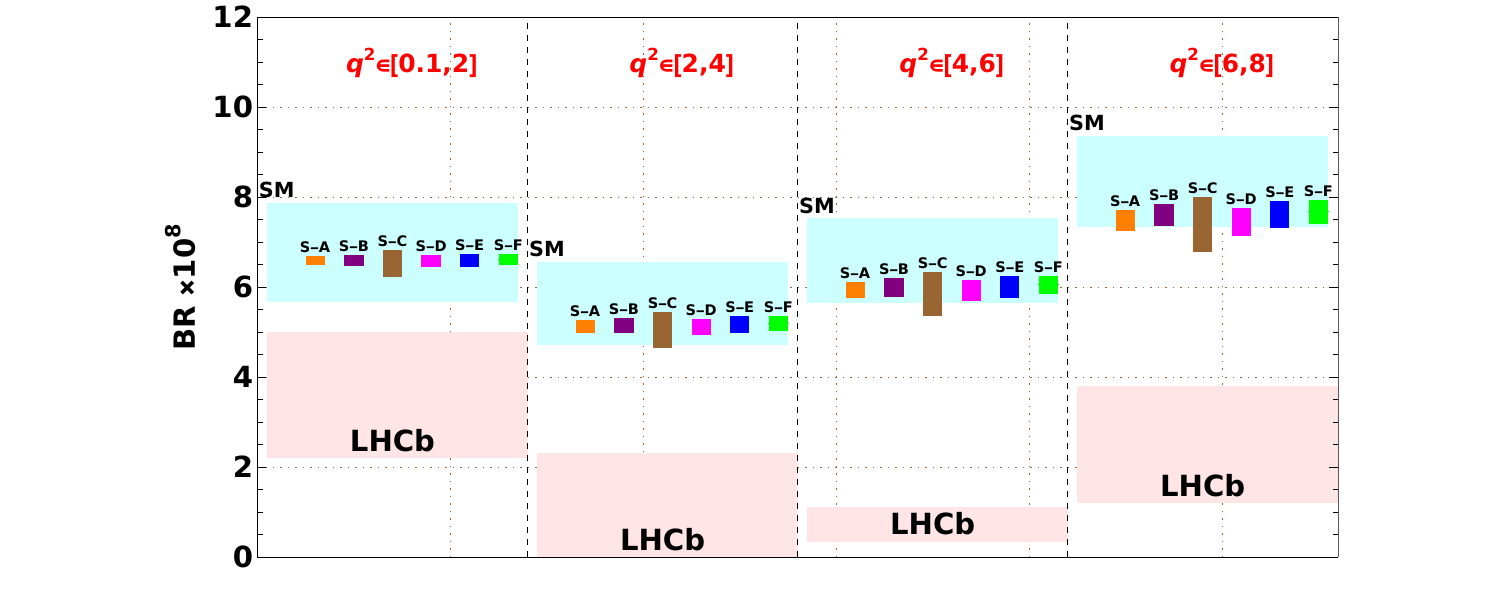}
    \hspace*{-1.7cm}\includegraphics[width=17.6cm,height=5.8cm]{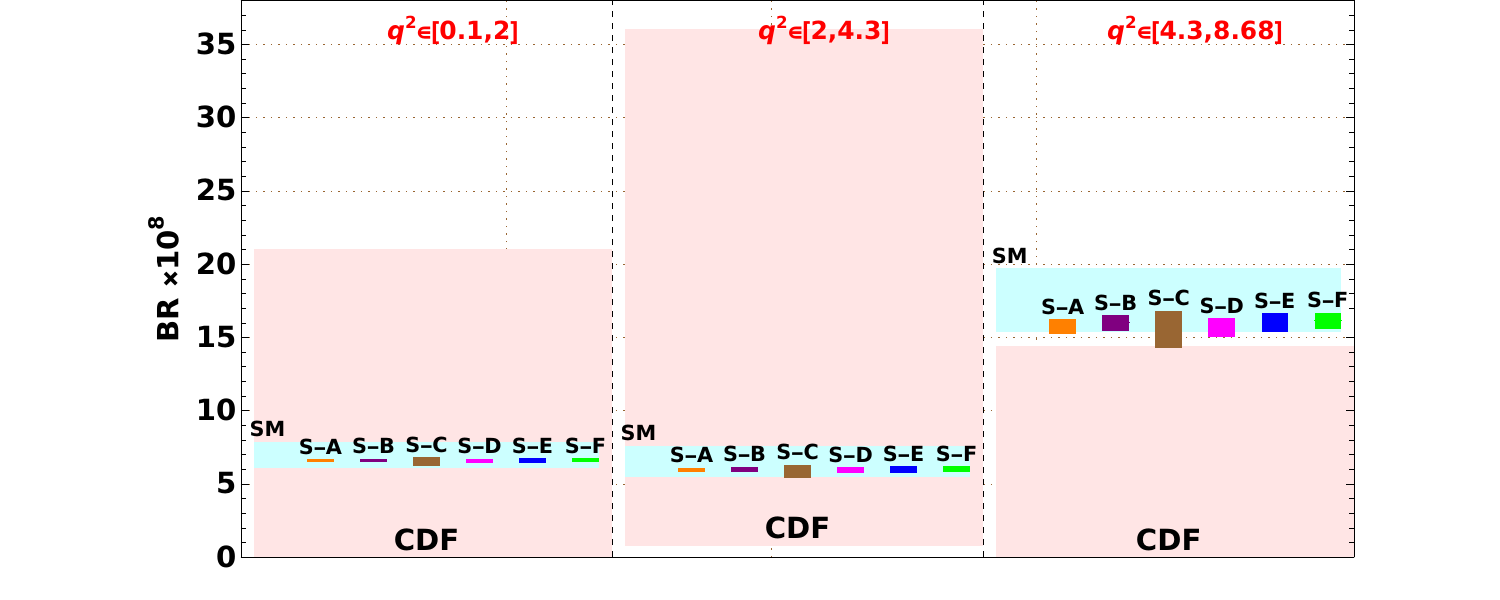}
    \caption{The branching ratios of the decay $\Lambda_b \to \Lambda \mu^+ \mu^-$ in the SM and various NP scenarios, shown across different $q^2$ bins. The experimental uncertainties from the HFLAV average (top panel), LHCb (middle panel), and CDF (bottom panel) are represented by light pink bands.}
    \label{Fig:BR-Exp-bin}
\end{figure}
Furthermore, we compare the binwise branching ratios of the $\Lambda_b$ baryonic decay individually with the available experimental measurements from LHCb, CDF, and the HFLAV average, as shown in Fig.~\ref{Fig:BR-Exp-bin}. In all plots, the cyan band represents the SM prediction, with its width denoting the $1\sigma$ theoretical uncertainty. The NP scenarios are distinguished by the following color scheme: orange for $\mathbf{S\text{-}A}$, purple for $\mathbf{S\text{-}B}$, brown for $\mathbf{S\text{-}C}$, magenta for $\mathbf{S\text{-}D}$, blue for $\mathbf{S\text{-}E}$, and green for $\mathbf{S\text{-}F}$. The shaded regions corresponding to each NP scenario indicate the $1\sigma$ predicted range arising from the associated Wilson coefficients. The experimental uncertainties are represented by light pink bands.  We observe that the predicted branching ratios in the presence of NP lie within the SM theoretical uncertainties in the first $q^2$ bin. Notably, the prediction in the $\mathbf{S\text{-}C}$ scenario, characterized by the NP couplings $(C_9^{\rm NP},\, C_{10}^{\rm NP})$, exhibits a visible deviation from the SM expectation and trends closer to the experimental data. However, due to the relatively large theoretical uncertainties in the branching ratio predictions of the $\Lambda_b \to \Lambda \mu^+ \mu^-$ decay, it remains challenging to draw definitive conclusions regarding the presence of NP from these measurements alone.

\begin{figure}[htp]
\flushleft
\vspace{-1cm}
\hspace*{-1.7cm}\includegraphics[width=18cm,height=6cm]{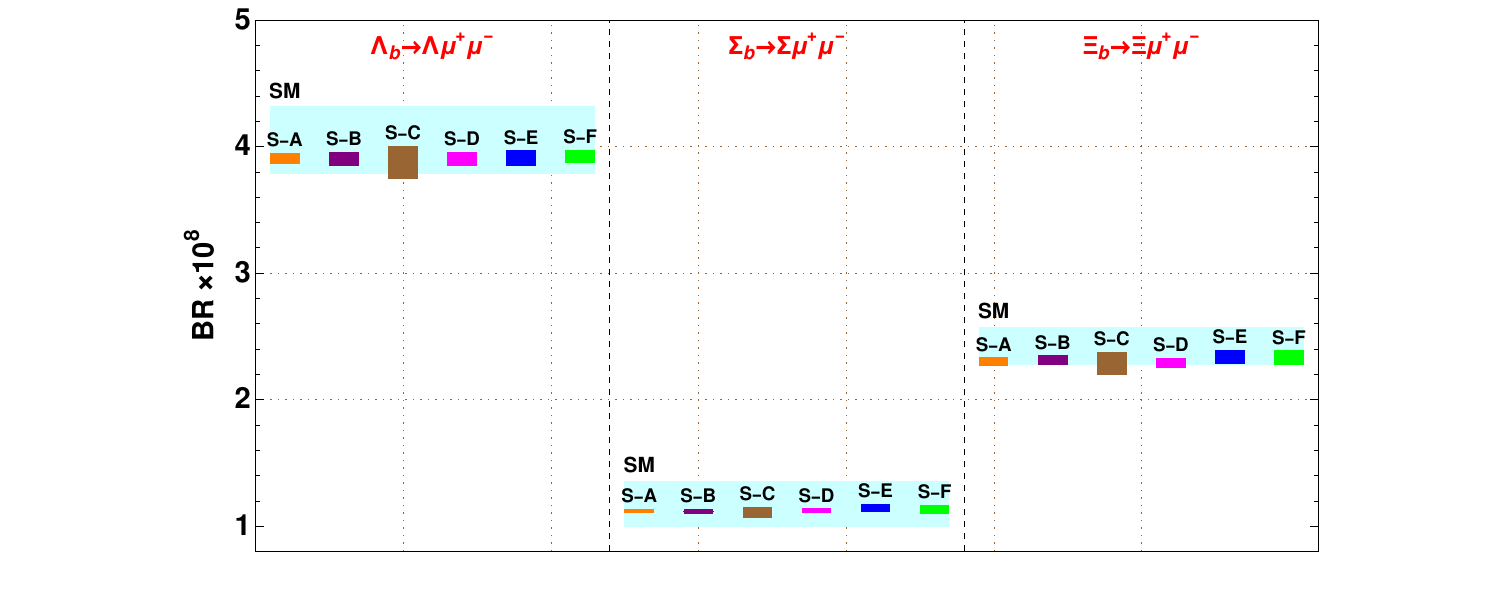}\\
\hspace*{-1.7cm}\includegraphics[width=18cm,height=6cm]{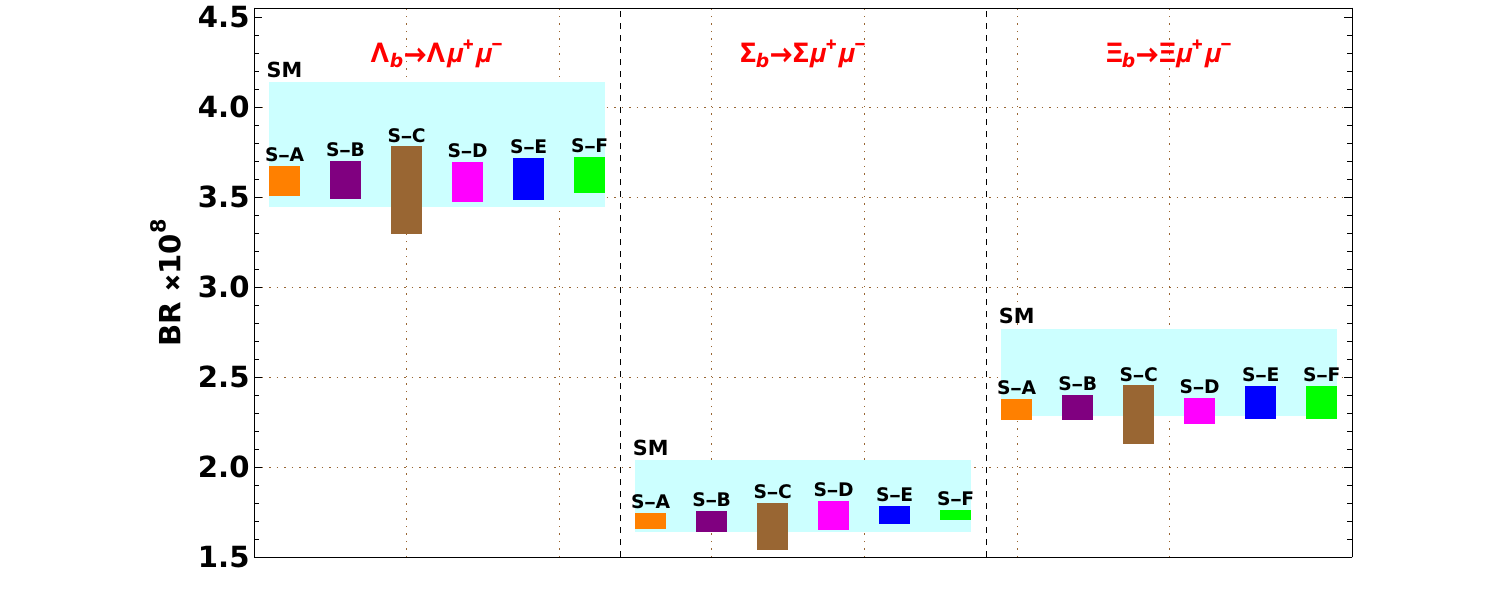}\\
\hspace*{-1.7cm}\includegraphics[width=18cm,height=6cm]{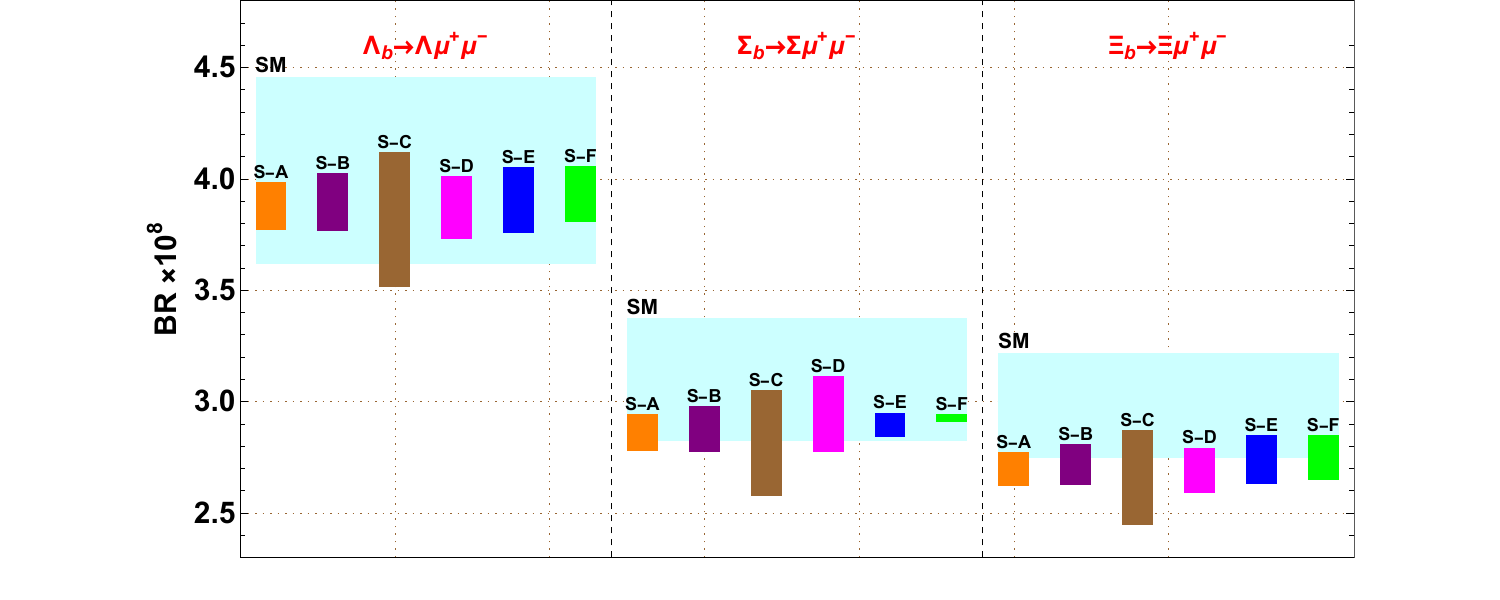}\\
\hspace*{-1.7cm}\includegraphics[width=18cm,height=6cm]{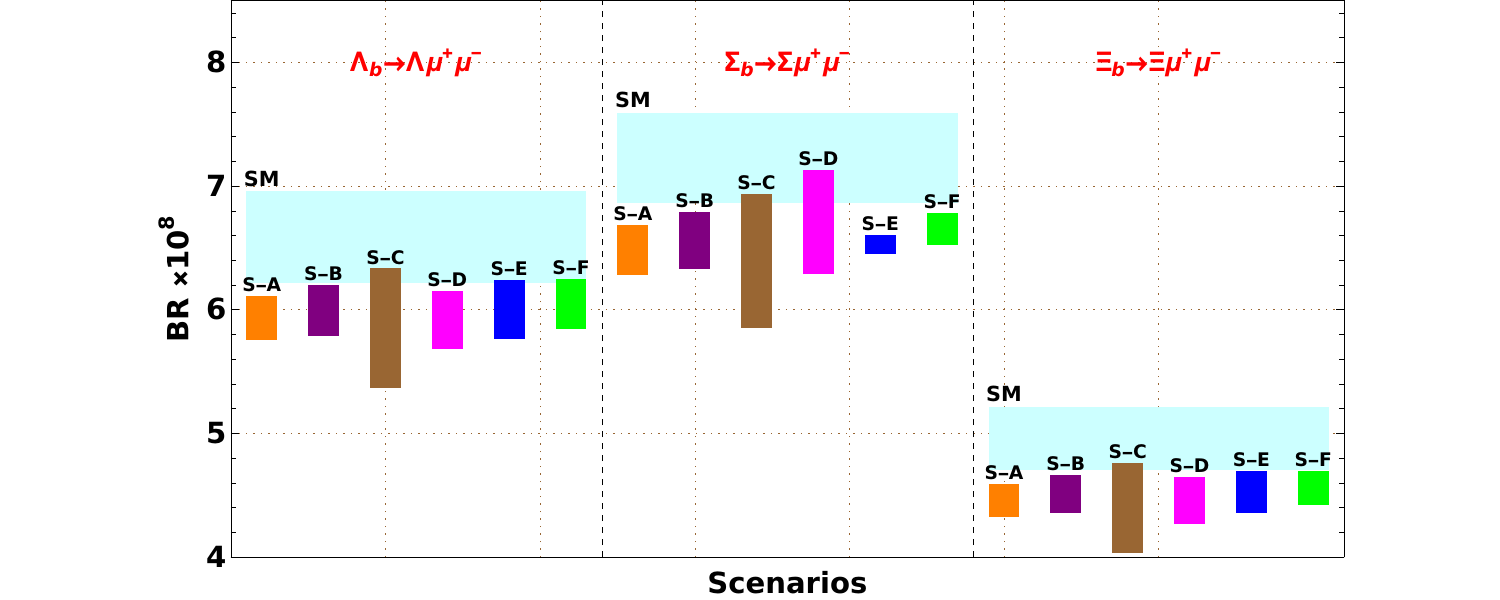}
\caption{$\mathcal{BR}(\Lambda_b \to \Lambda\mu \mu)$ (left), $\mathcal{BR}(\Sigma_b \to \Sigma\mu \mu)$ (middle) and $\mathcal{BR}(\Xi_b \to \Xi \mu\mu$) (right) are shown across $q^2$ bins: $q^2 \in[0.1,0.98]$ (top), $q^2 \in[1.1,2.5]$ (second from top), $q^2 \in[2.5,4.0]$ (second from bottom), $q^2 \in[4.0,6.0]$ (bottom). The bands indicate $1\sigma$ uncertainty.}
\label{Fig:BR}
\end{figure}

\begin{figure}[htp]
\flushleft
\vspace{-0.8cm}
\hspace*{-1.7cm}\includegraphics[width=18cm,height=6cm]{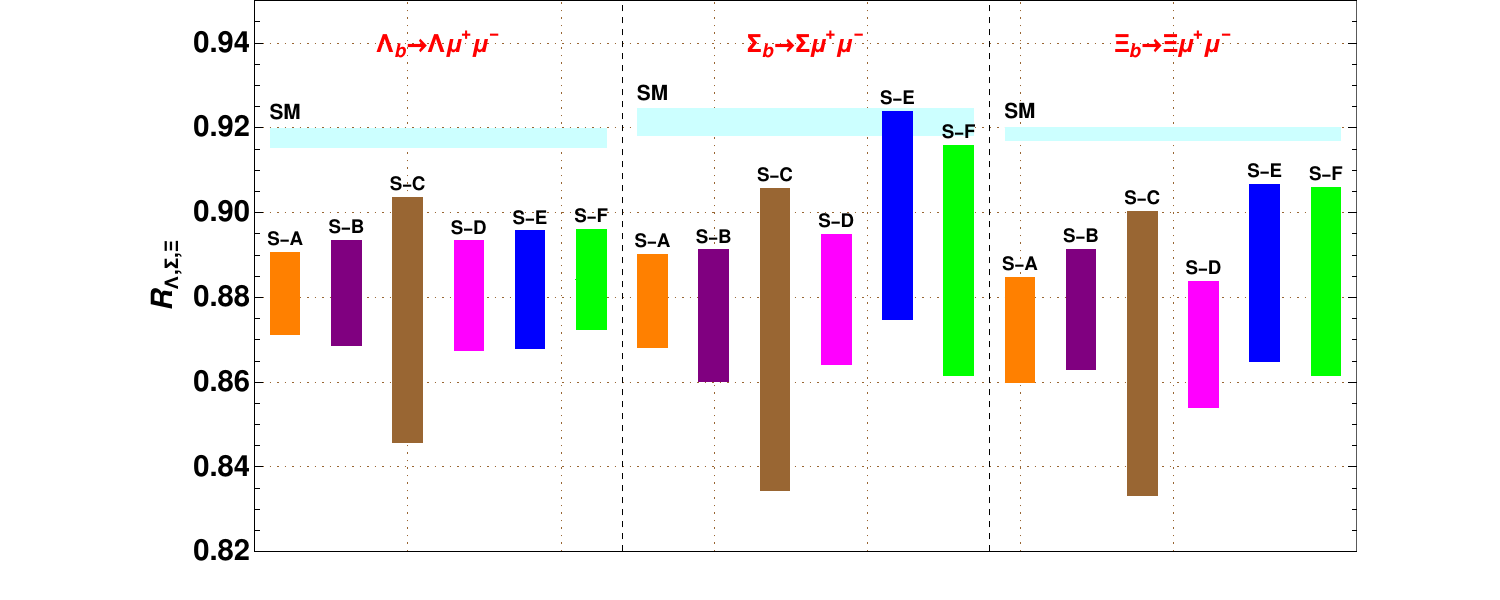}\\
\hspace*{-1.7cm}\includegraphics[width=18cm,height=6cm]{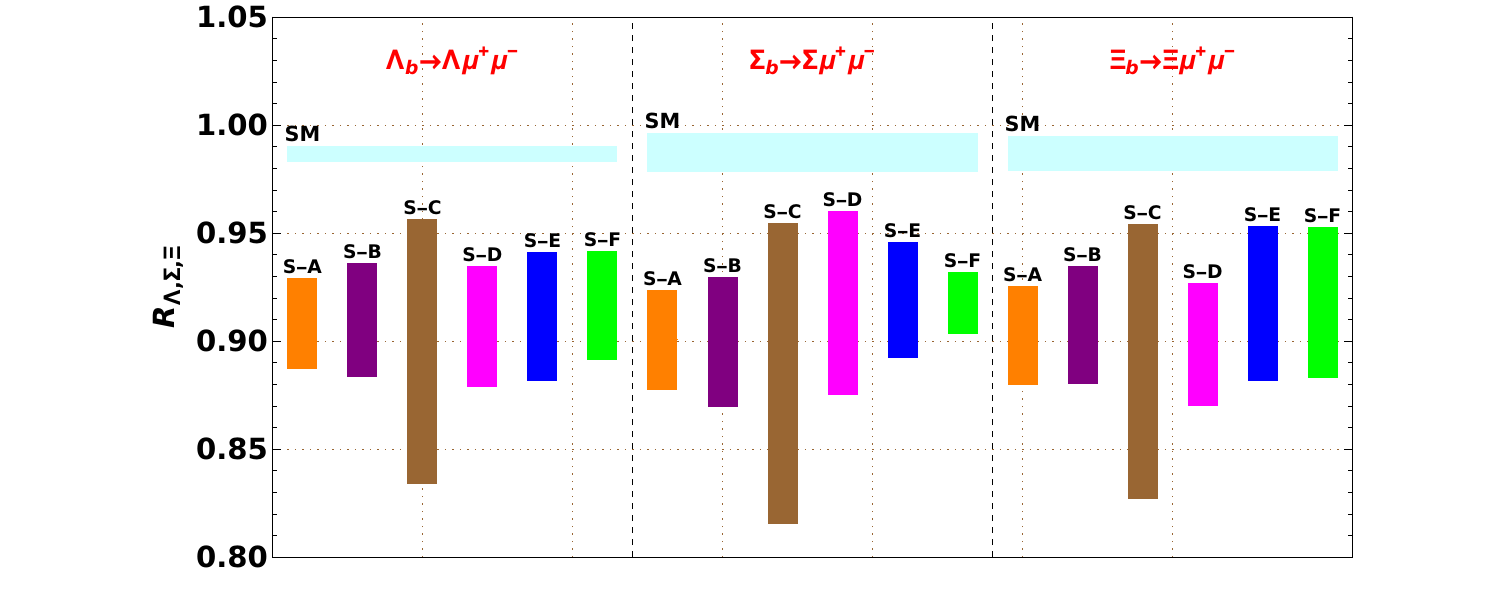}\\
\hspace*{-1.7cm}\includegraphics[width=18cm,height=6cm]{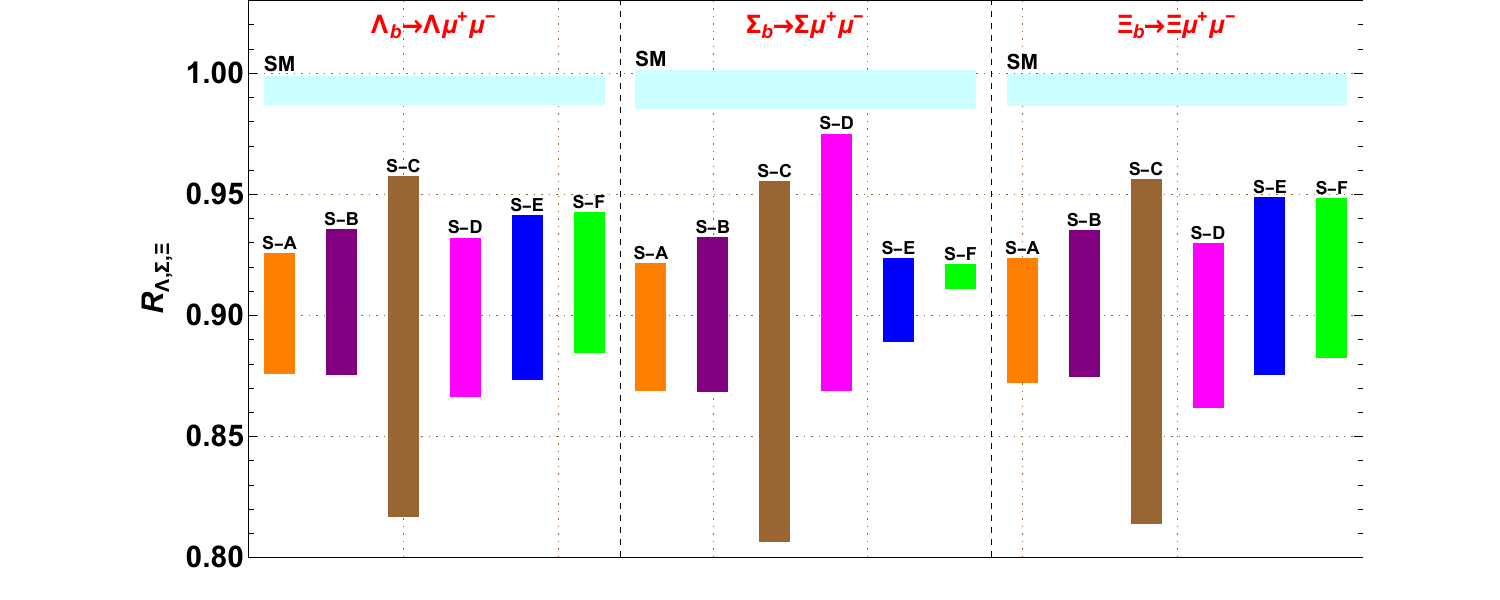}\\
\hspace*{-1.7cm}\includegraphics[width=18cm,height=6cm]{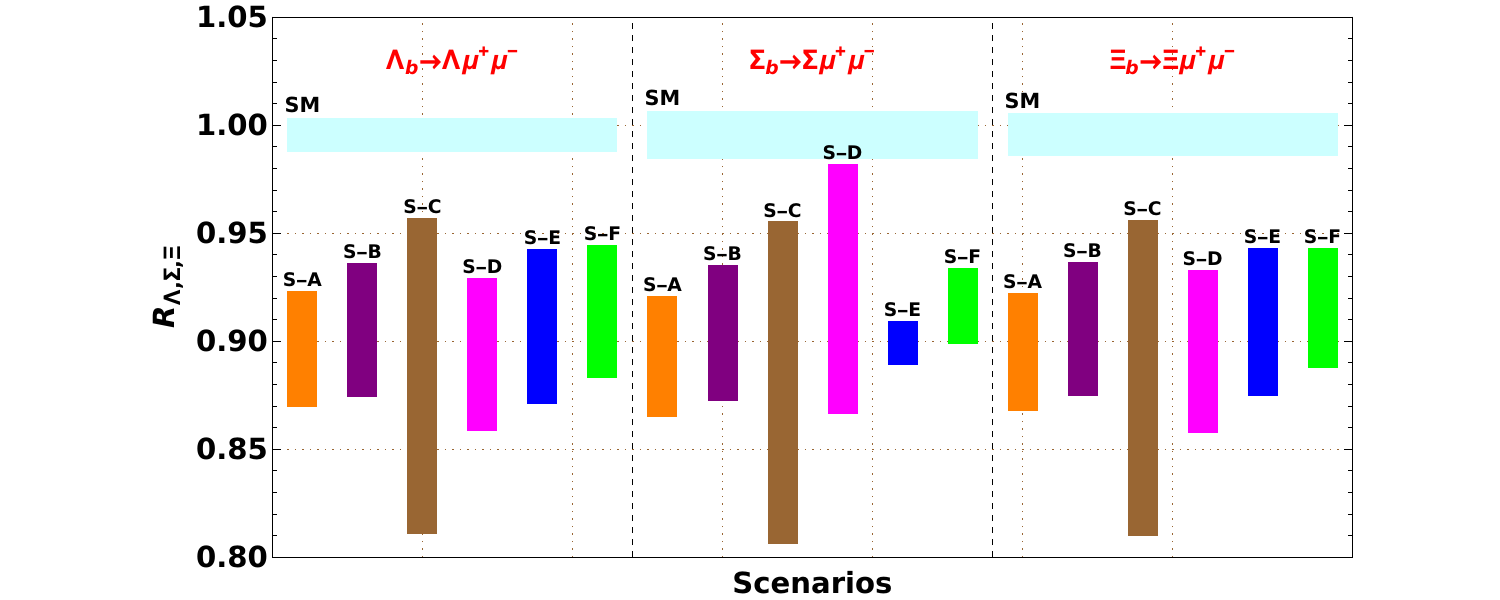}
\caption{Same as Fig. \ref{Fig:BR}, illustrating the observables $\mathcal{R}_{\Lambda}$ (left), $\mathcal{R}_{\Sigma}$ (middle), and $\mathcal{R}_{\Xi}$ (right).}
\label{Fig:LNU}
\end{figure}

\begin{figure}[htp]
\flushleft
\vspace{-0.8cm}
\hspace*{-1.7cm}\includegraphics[width=18cm,height=6cm]{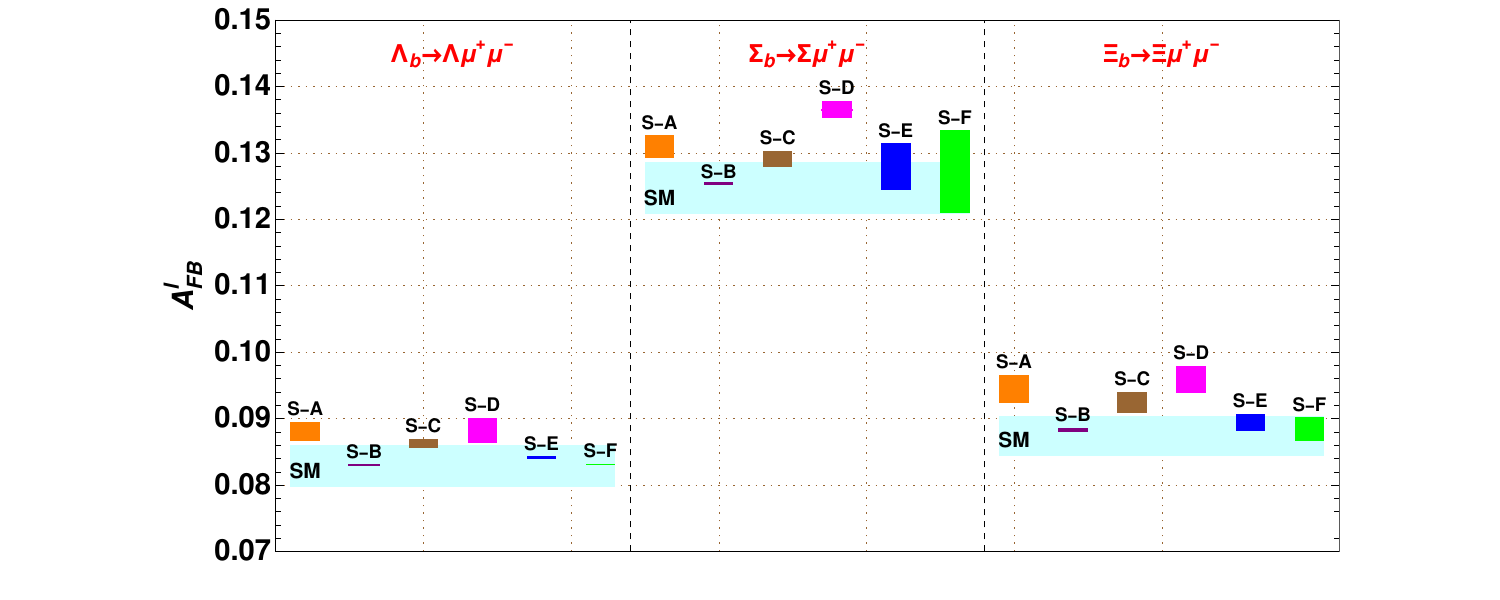}\\
\hspace*{-1.7cm}\includegraphics[width=18cm,height=6cm]{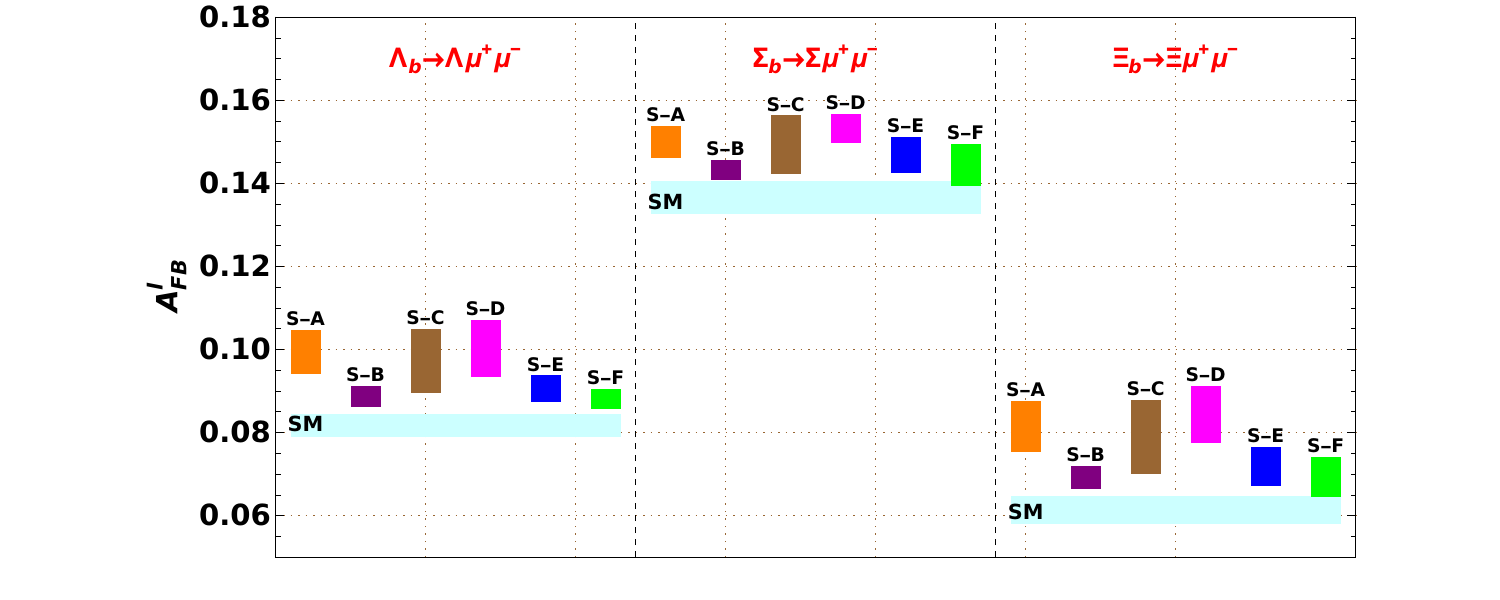}\\
\hspace*{-1.7cm}\includegraphics[width=18cm,height=6cm]{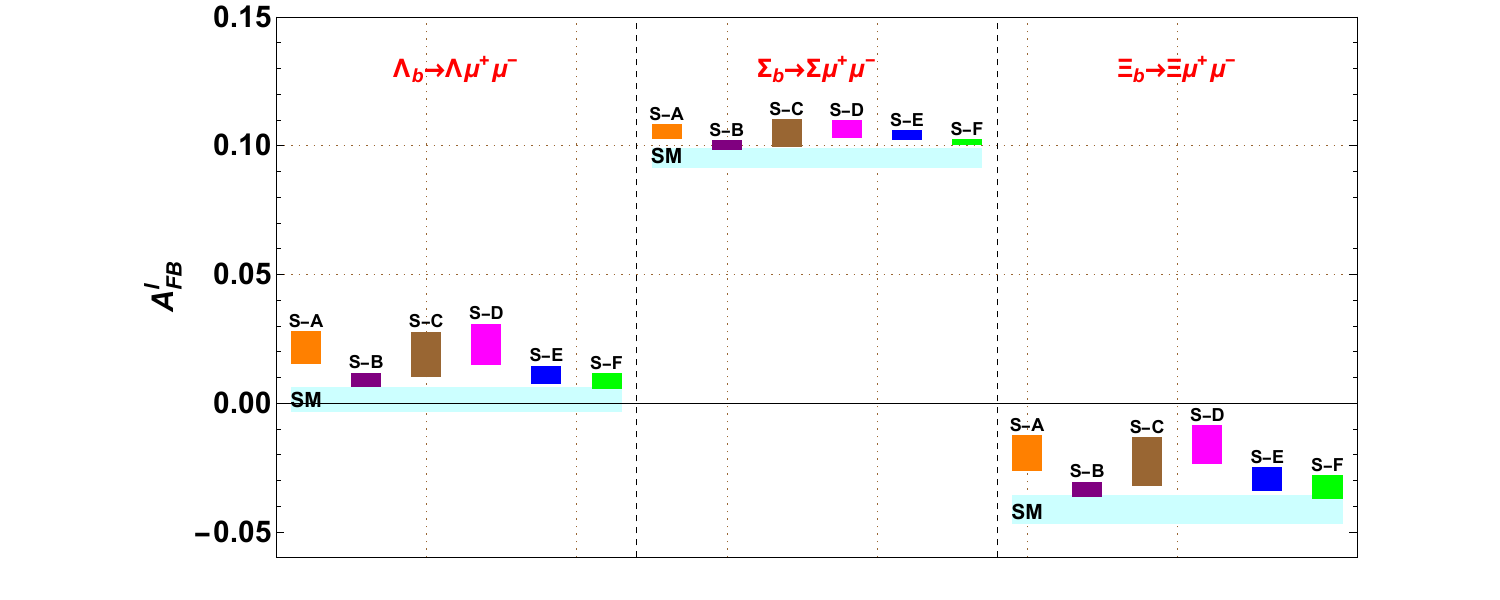}\\
\hspace*{-1.7cm}\includegraphics[width=18cm,height=6cm]{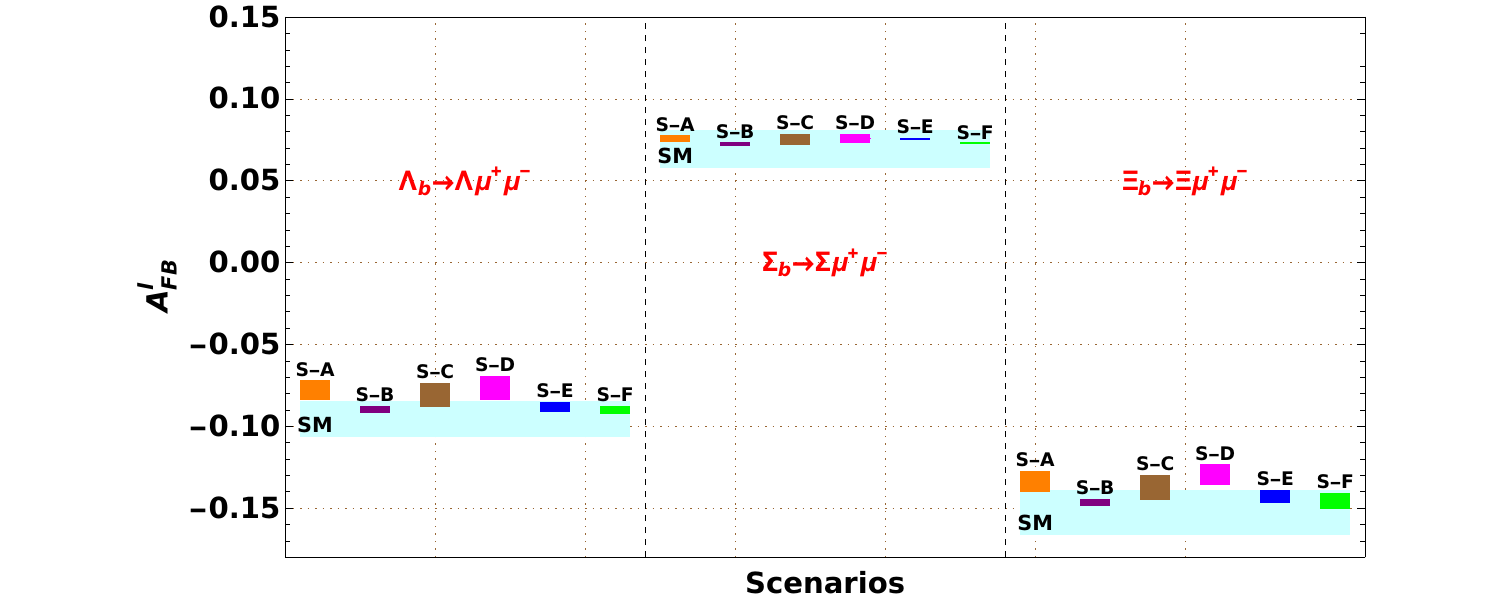}
\caption{Same as Fig. \ref{Fig:BR}, focusing the observables $\mathcal{A}_{FB}^\ell(\Lambda_b \to \Lambda\mu \mu)$ (left), $\mathcal{A}_{FB}^\ell(\Sigma_b \to \Sigma\mu \mu)$ (middle) and $\mathcal{A}_{FB}^\ell (\Xi_b \to \Xi \mu\mu$) (right).}
\label{Fig:AFBlep}
\end{figure}

\begin{figure}[htp]
\flushleft
\vspace{-0.8cm}
\hspace*{-1.7cm}\includegraphics[width=18cm,height=6cm]{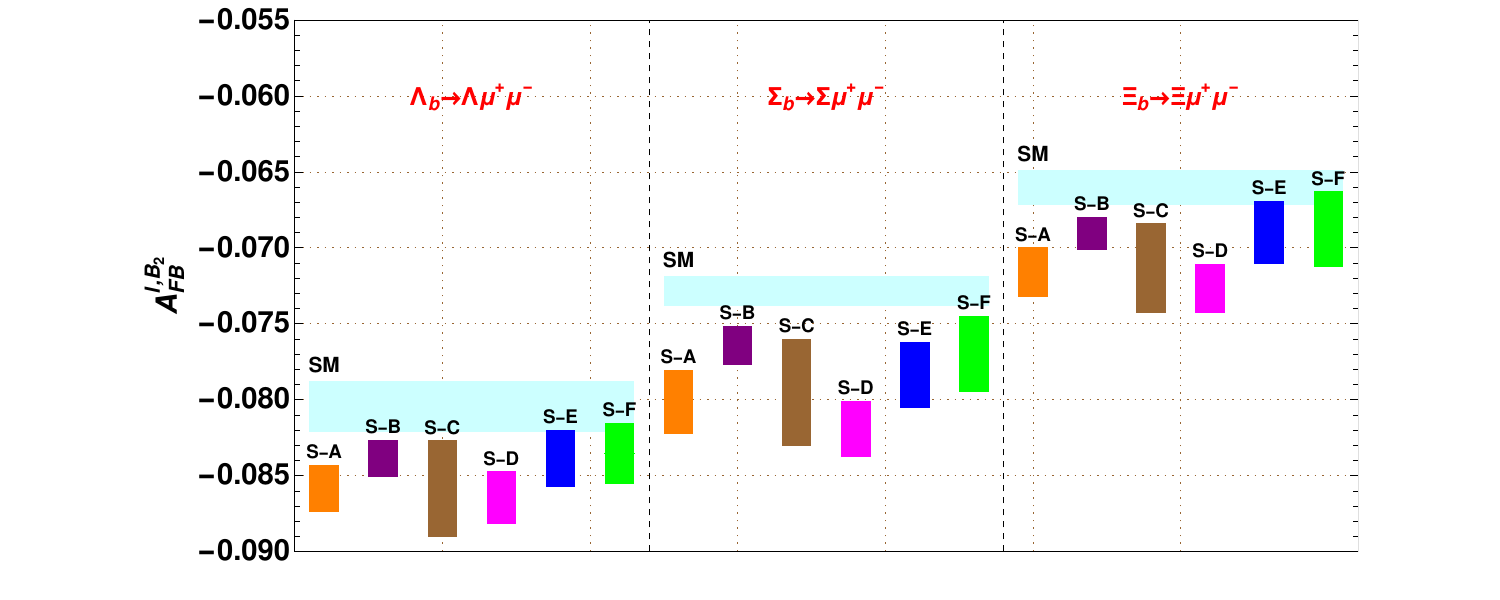}\\
\hspace*{-1.7cm}\includegraphics[width=18cm,height=6cm]{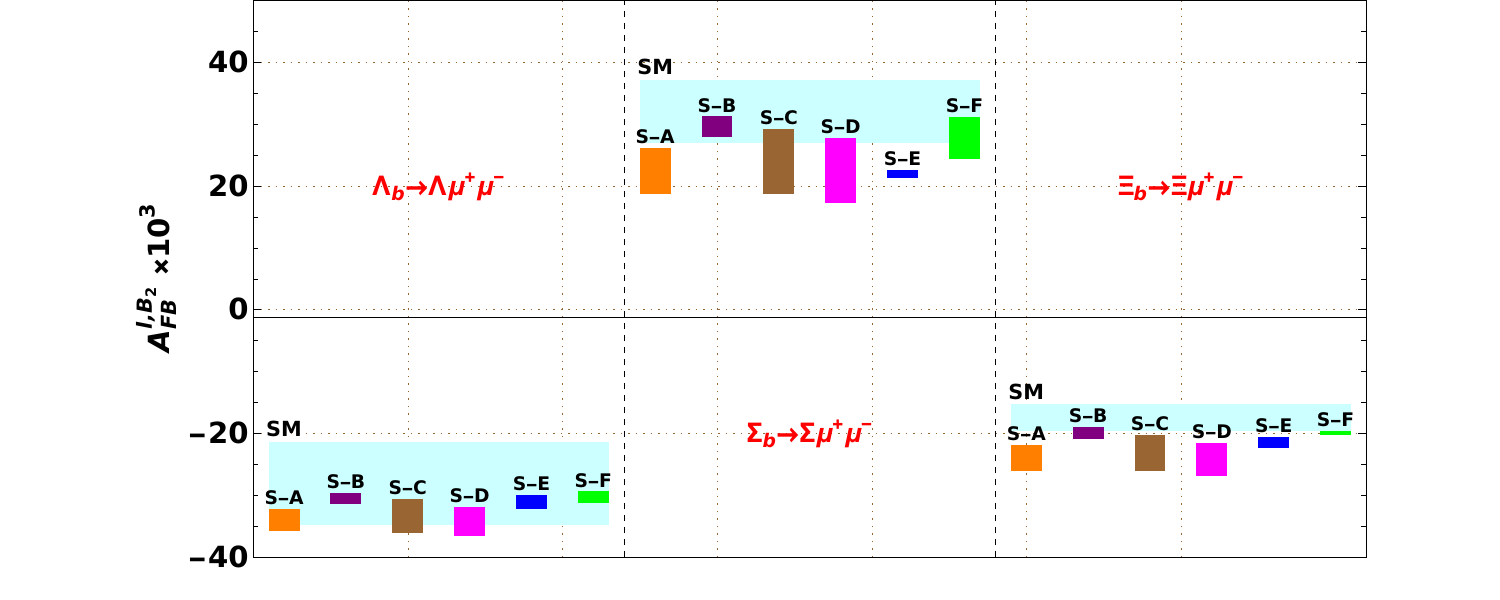}\\
\hspace*{-1.7cm}\includegraphics[width=18cm,height=6cm]{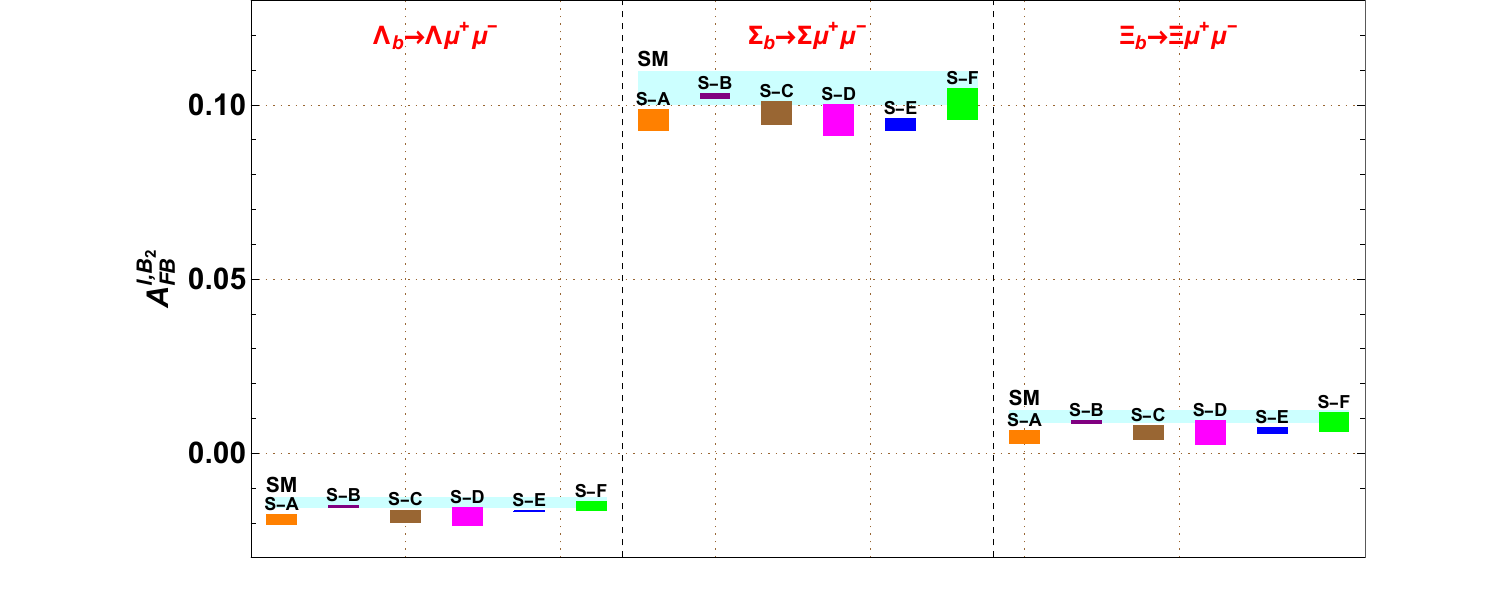}\\
\hspace*{-1.7cm}\includegraphics[width=18cm,height=6cm]{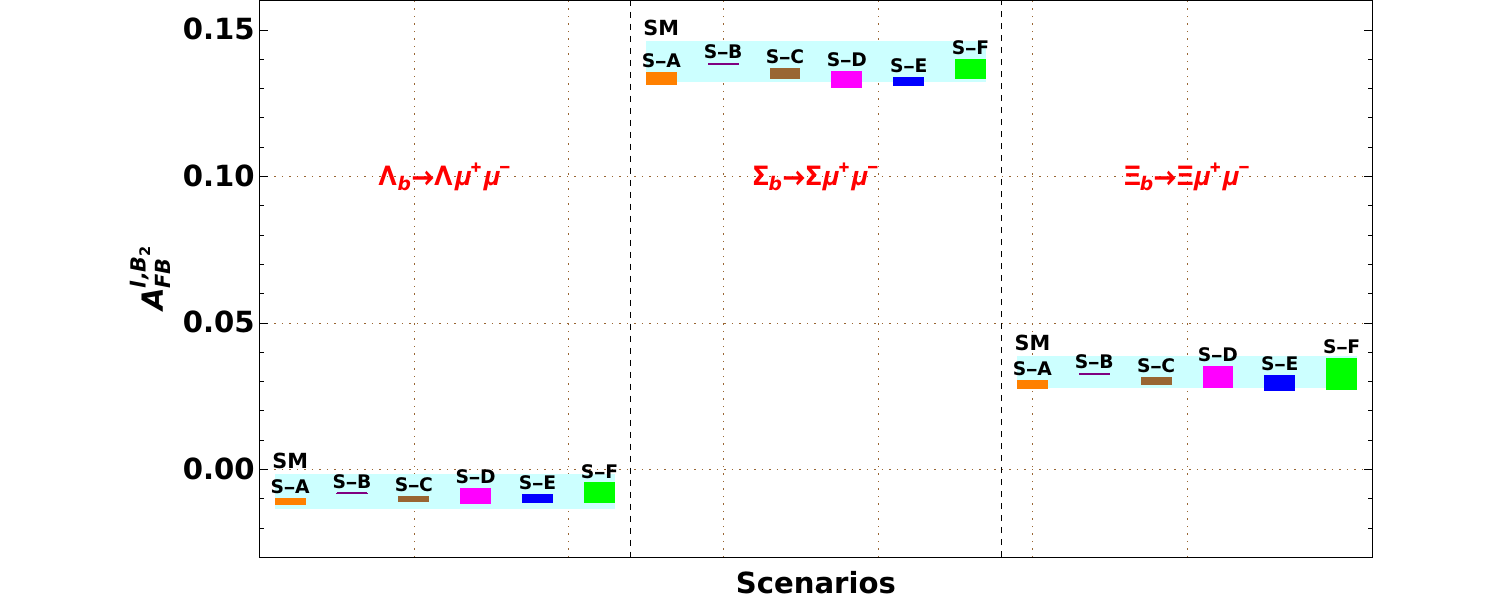}
\caption{Same as Fig. \ref{Fig:BR}, displaying the observables $\mathcal{A}_{FB}^{\ell, B_2}(\Lambda_b \to \Lambda\mu \mu)$ (left), $\mathcal{A}_{FB}^{\ell, B_2}(\Sigma_b \to \Sigma\mu \mu)$ (middle) and $\mathcal{A}_{FB}^{\ell, B_2} (\Xi_b \to \Xi \mu\mu$) (right).}
\label{Fig:AFBlephad}
\end{figure}

\begin{figure}[htp]
\flushleft
\vspace{-0.8cm}
\hspace*{-1.7cm}\includegraphics[width=18cm,height=6cm]{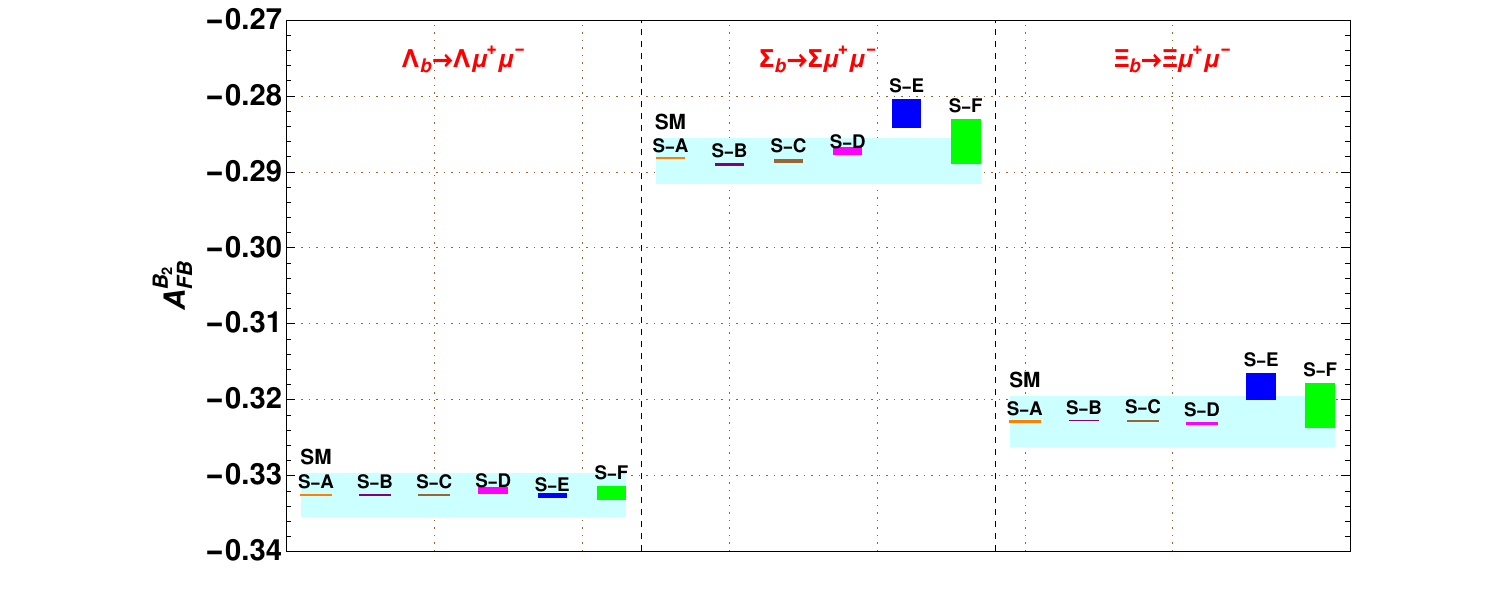}\\
\hspace*{-1.7cm}\includegraphics[width=18cm,height=6cm]{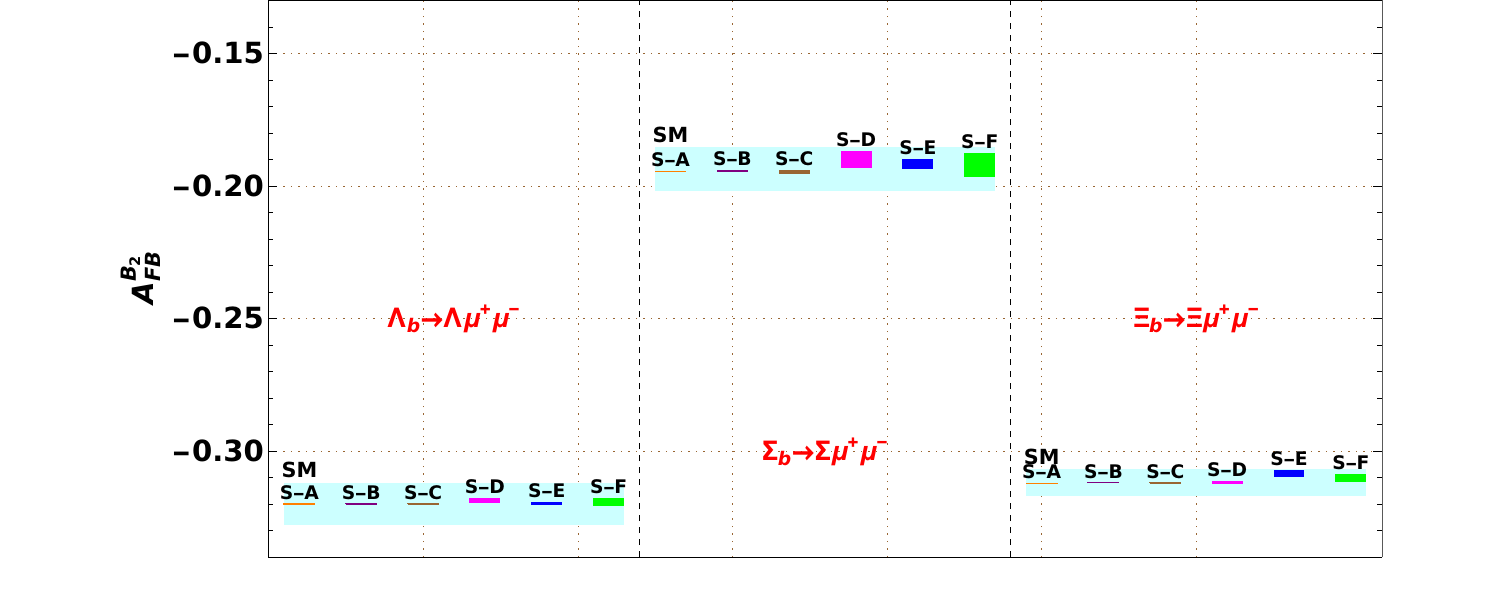}\\
\hspace*{-1.7cm}\includegraphics[width=18cm,height=6cm]{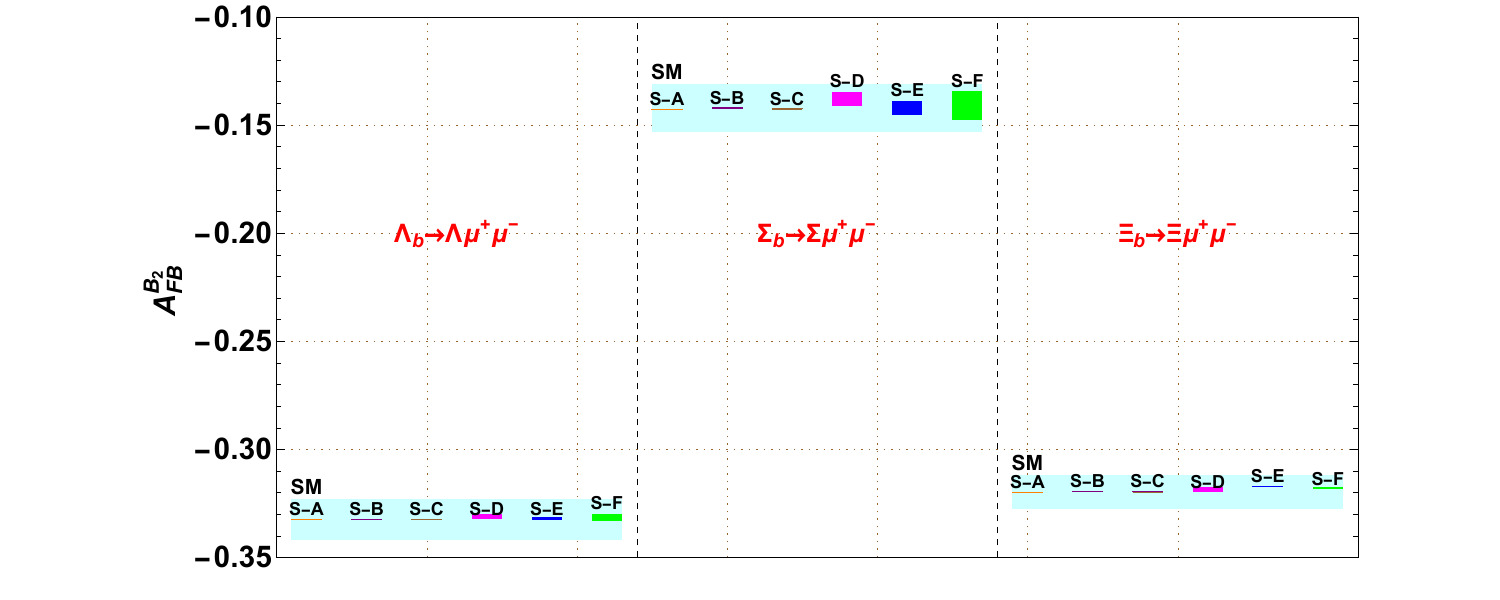}\\
\hspace*{-1.7cm}\includegraphics[width=18cm,height=6cm]{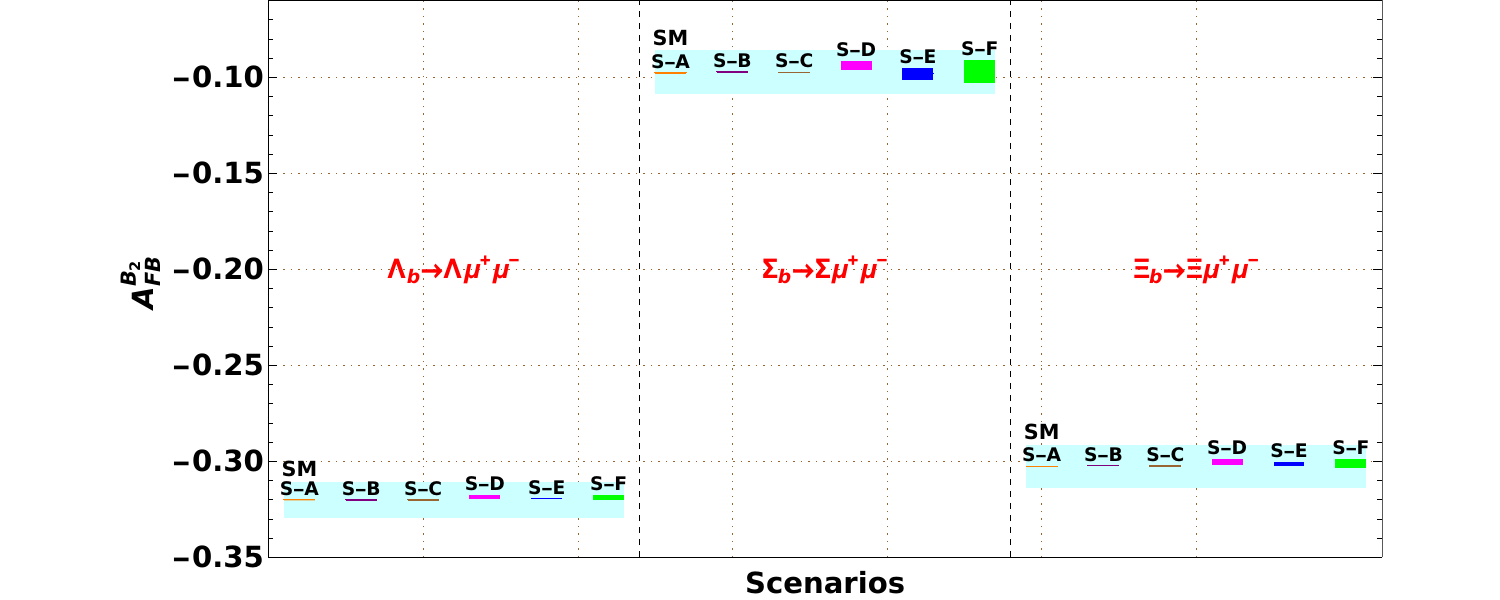}
\caption{Same as Fig. \ref{Fig:BR}, emphasizing the observables $\mathcal{A}_{FB}^{B_2}(\Lambda_b \to \Lambda\mu \mu)$ (left), $\mathcal{A}_{FB}^{B_2}(\Sigma_b \to \Sigma\mu \mu)$ (middle) and $\mathcal{A}_{FB}^{B_2} (\Xi_b \to \Xi \mu\mu$) (right).}
\label{Fig:AFBhad}
\end{figure}

\begin{figure}[htp]
\flushleft
\vspace{-0.8cm}
\hspace*{-1.7cm}\includegraphics[width=18cm,height=6cm]{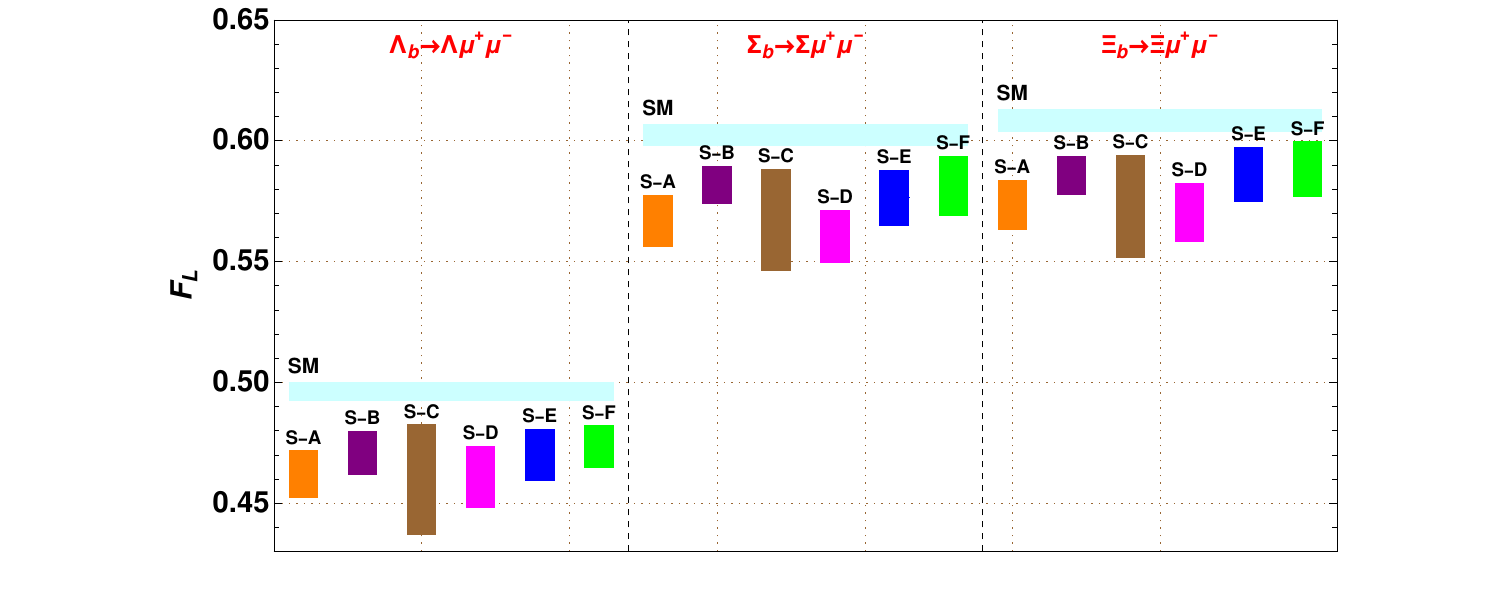}\\
\hspace*{-1.7cm}\includegraphics[width=18cm,height=6cm]{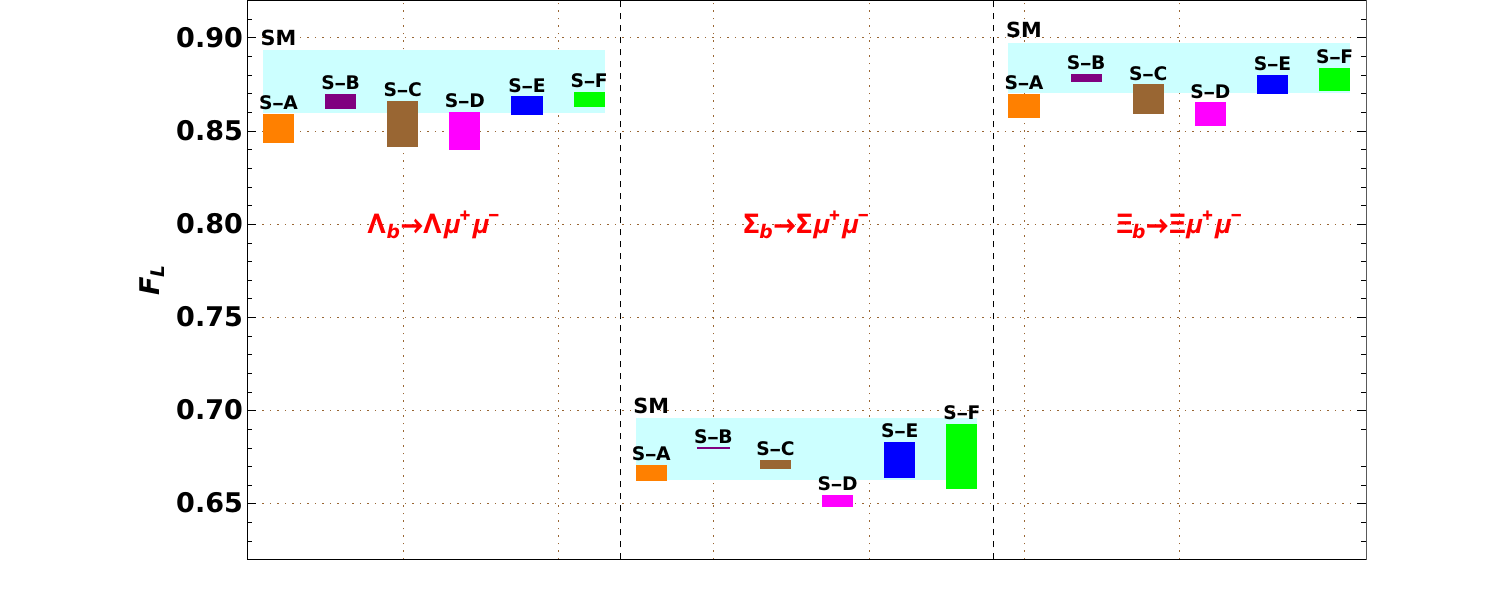}\\
\hspace*{-1.7cm}\includegraphics[width=18cm,height=6cm]{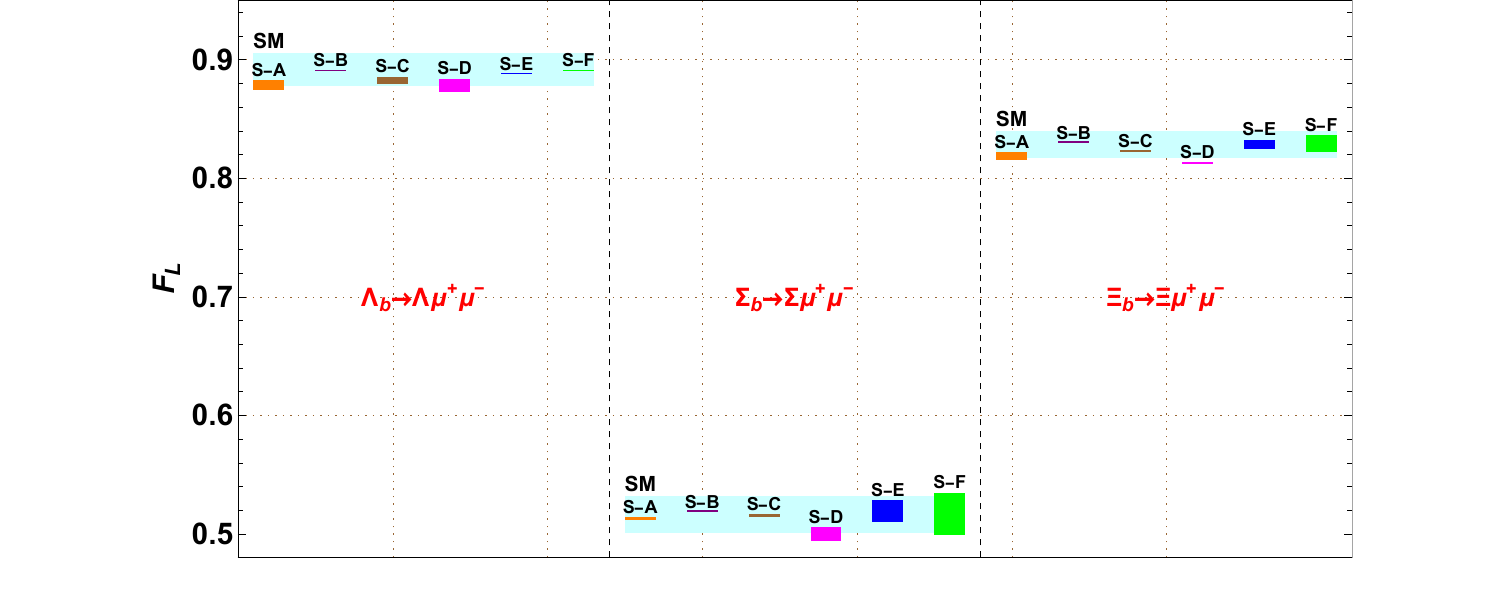}\\
\hspace*{-1.7cm}\includegraphics[width=18cm,height=6cm]{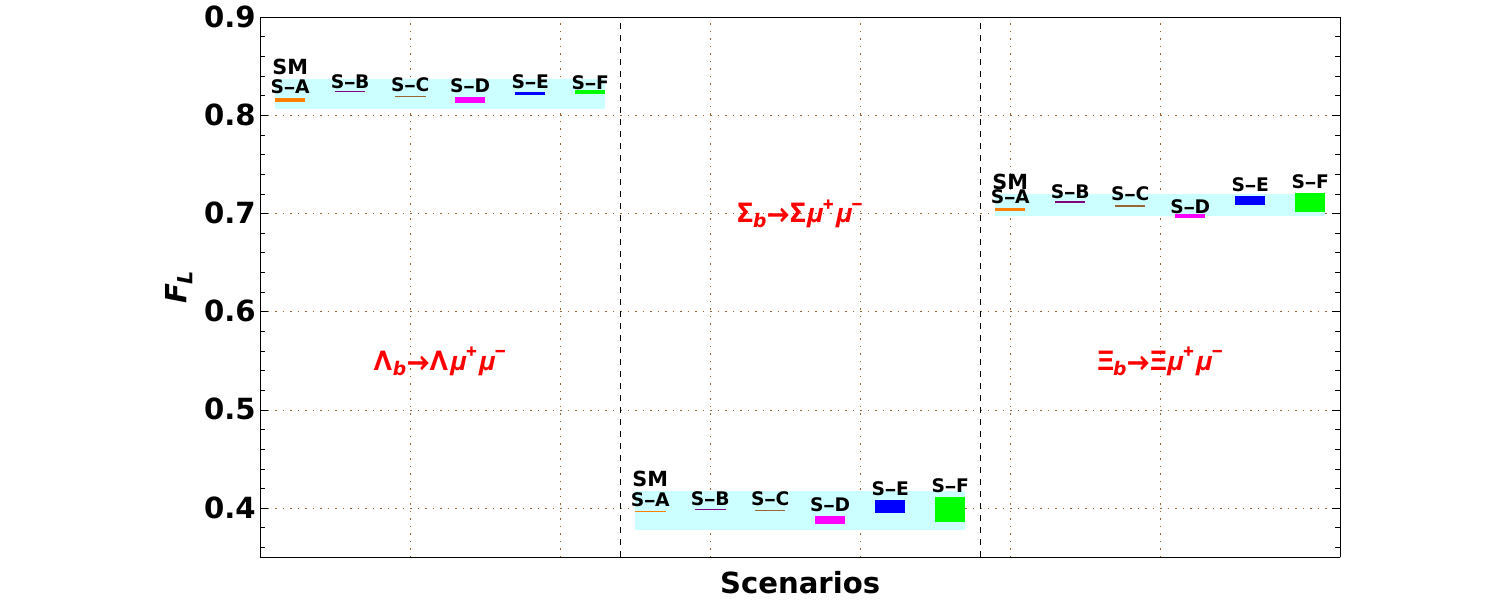}
\caption{Same as Fig. \ref{Fig:BR}, featuring the observables $\mathcal{F}_{L}(\Lambda_b \to \Lambda\mu \mu)$ (left), $\mathcal{F}_{L}(\Sigma_b \to \Sigma\mu \mu)$ (middle) and $\mathcal{F}_{L} (\Xi_b \to \Xi \mu\mu$) (right).}
\label{Fig:polarization}
\end{figure}

\begin{figure}[htp]
\flushleft
\vspace{-0.8cm}
\hspace*{-1.7cm}\includegraphics[width=18cm,height=6cm]{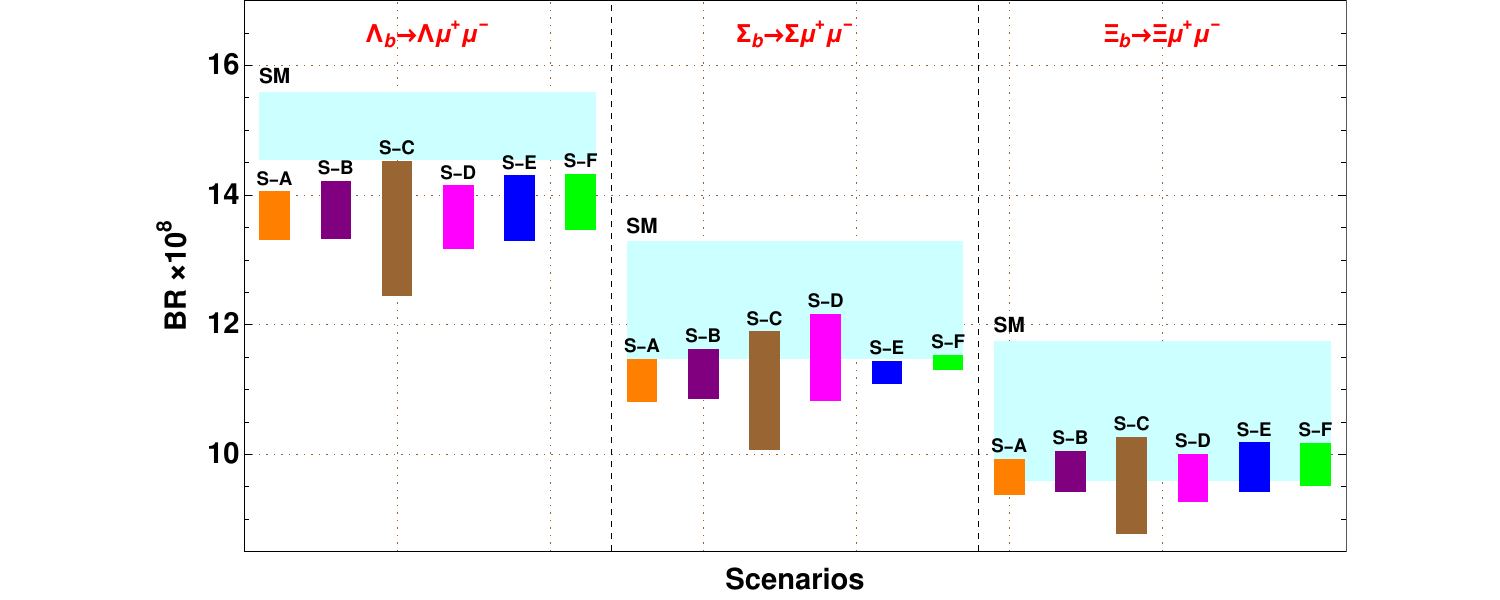}\\
\hspace*{-1.7cm}\includegraphics[width=18cm,height=6cm]{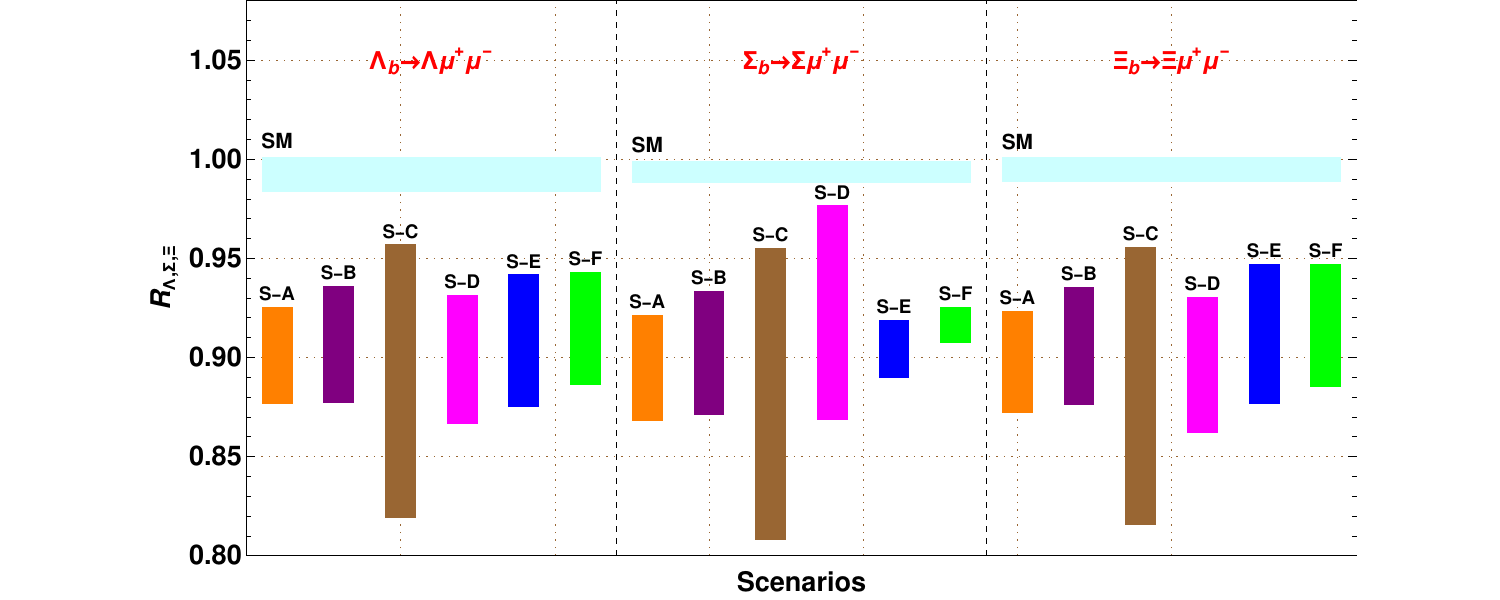}\\
\hspace*{-1.7cm}\includegraphics[width=18cm,height=6cm]{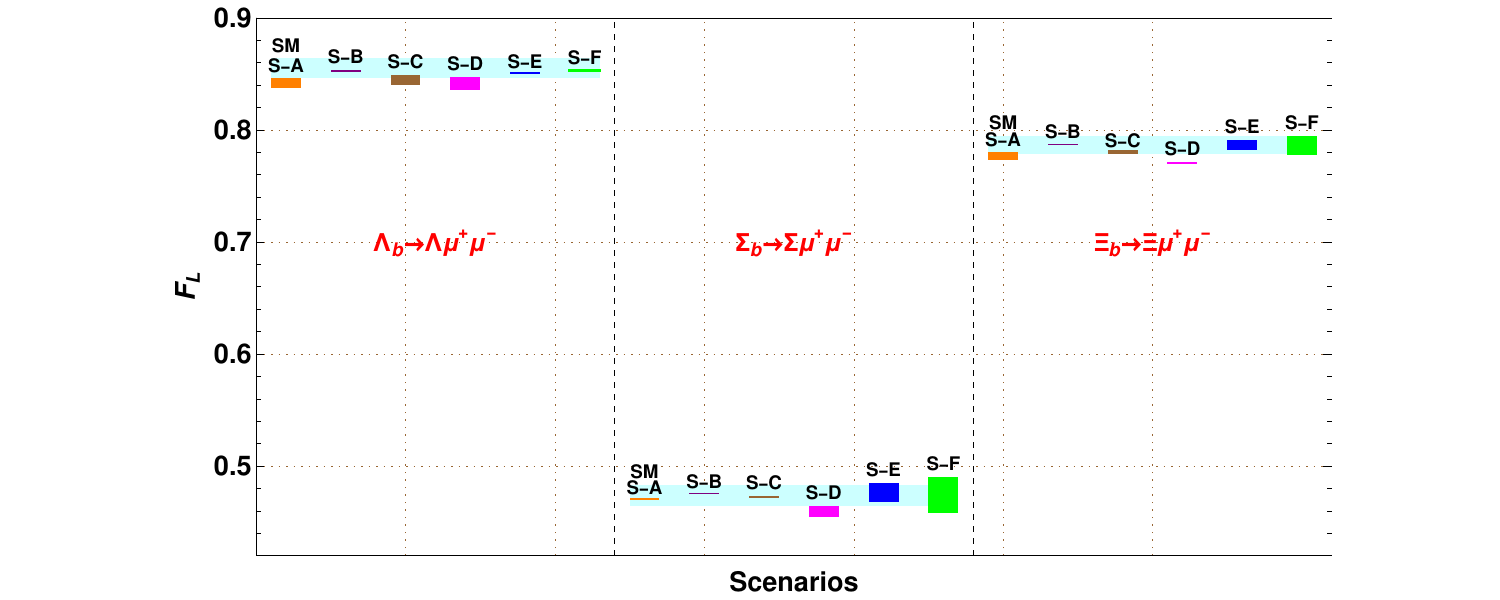}
\caption{Same as Fig. \ref{Fig:BR}, showcasing the observables: branching ratio (top), LNU parameter (middle), longitudinal polarisation asymmetry (bottom) in the low $q^2$ region $[1,6]~{\rm GeV}^2$.}
\label{Fig:BR1to6}
\end{figure}

\begin{figure}[htp]
\flushleft
\vspace{-0.8cm}
\hspace*{-1.7cm}\includegraphics[width=18cm,height=6cm]{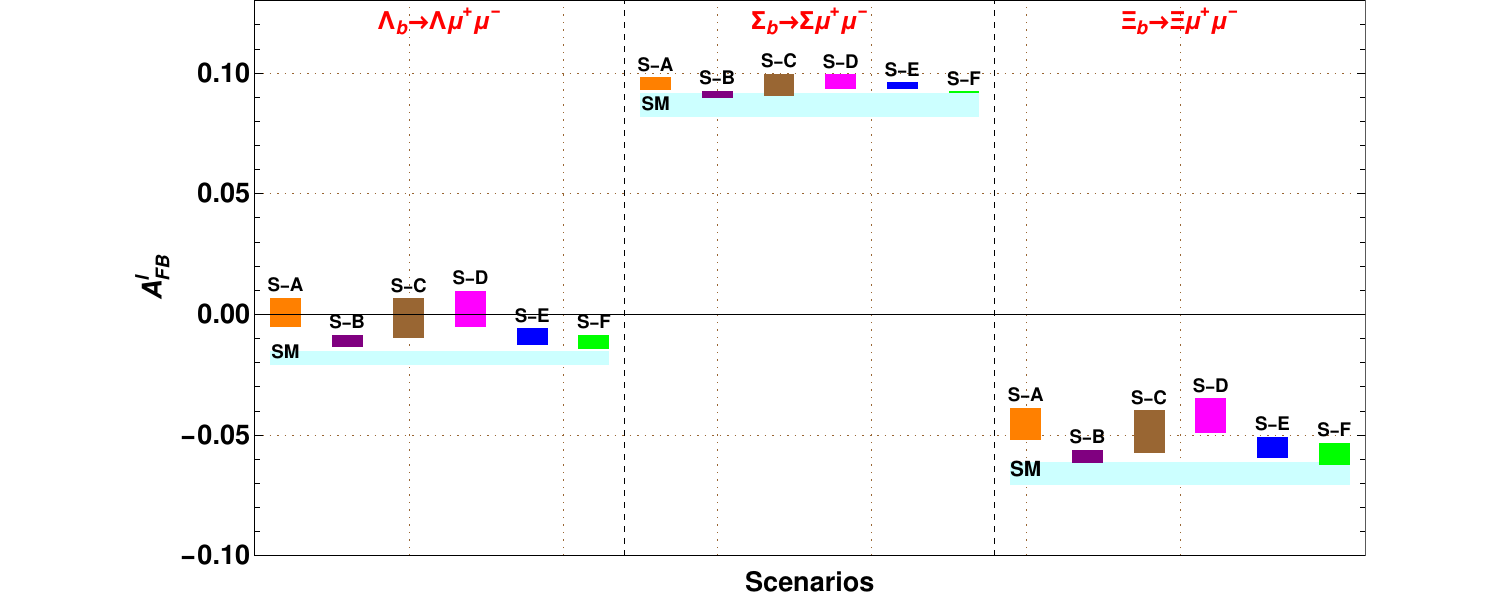}\\
\hspace*{-1.7cm}\includegraphics[width=18cm,height=6cm]{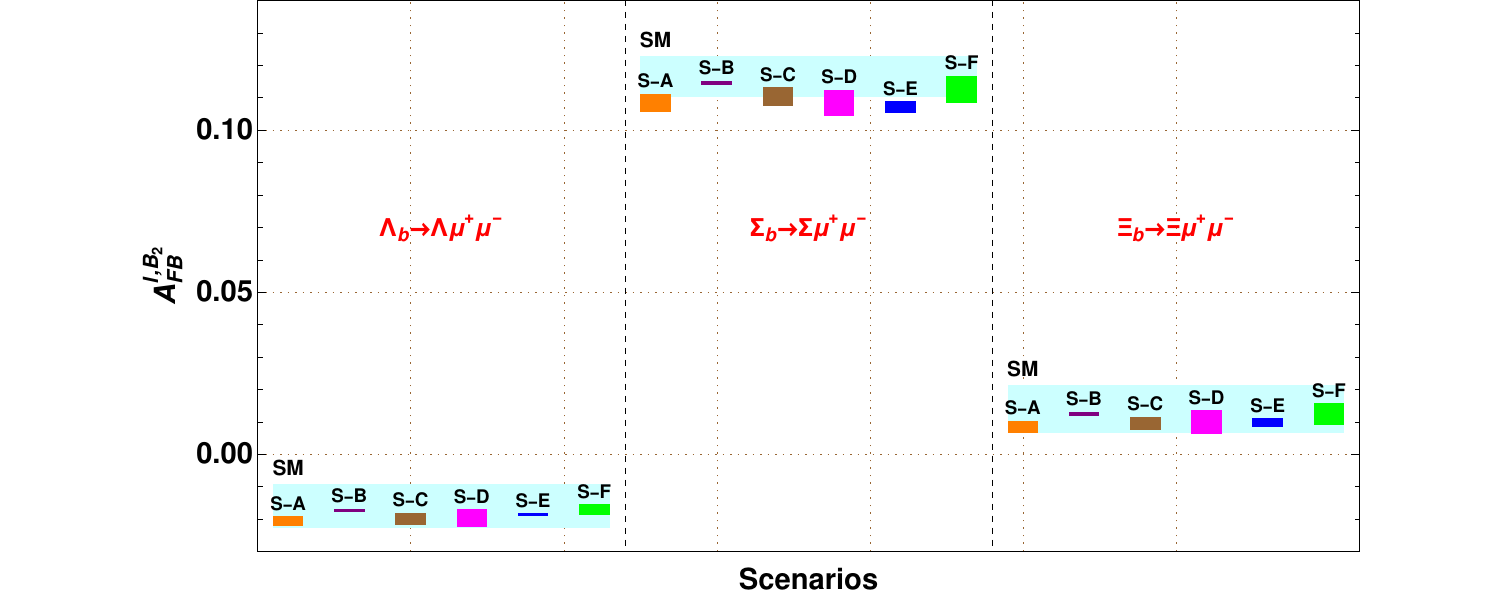}\\
\hspace*{-1.7cm}\includegraphics[width=18cm,height=6cm]{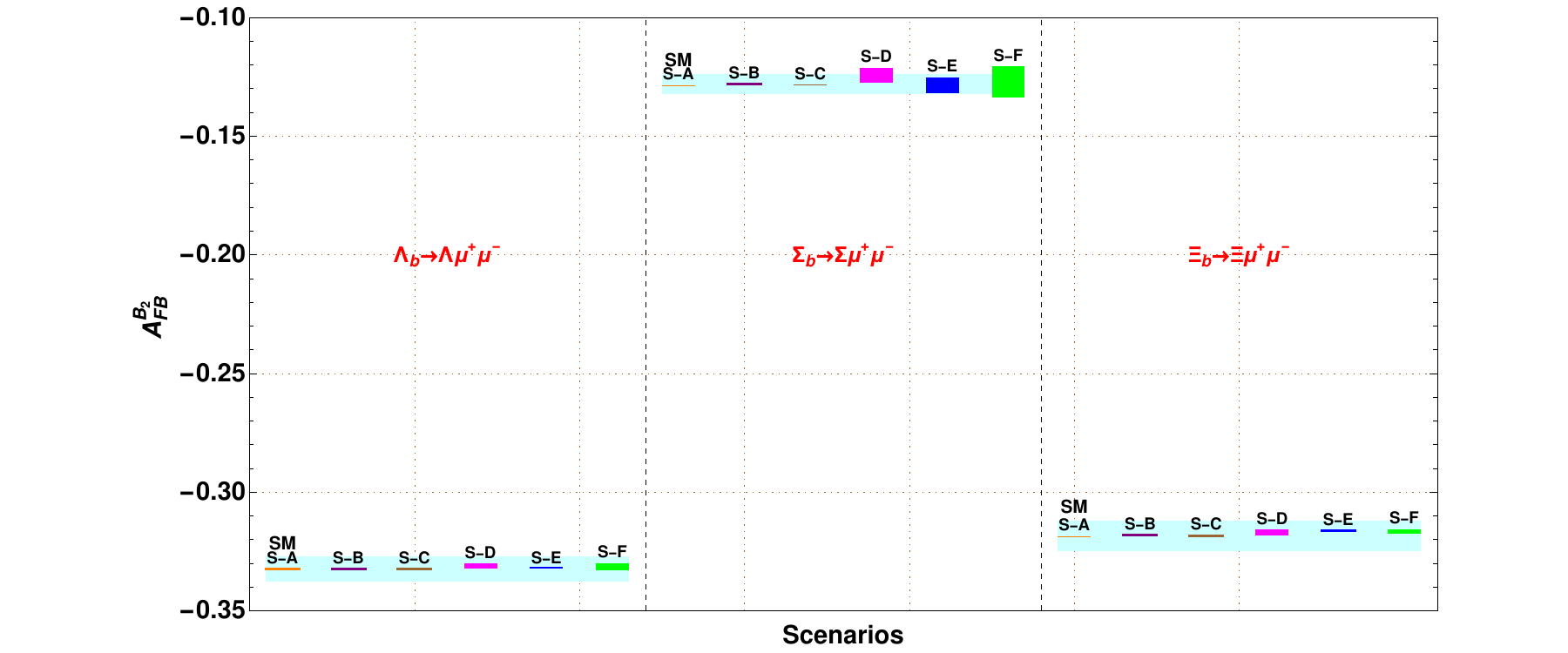}
\caption{Same as Fig. \ref{Fig:BR}, presenting the observables: $\mathcal{A}_{FB}^{\mu}$ (top), $\mathcal{A}_{FB}^{\mu, B_2}$ (middle) and $\mathcal{A}_{FB}^{B_2}$ (bottom) in the low $q^2 \in [1,6]~{\rm GeV}^2$ region.}
\label{Fig:AFB1to6}
\end{figure}

To further test the consistency and universality of the proposed NP scenarios, we extend our analysis to include related semileptonic baryonic decay modes, namely $\Sigma_b \to \Sigma \mu^+ \mu^-$ and $\Xi_b \to \Xi \mu^+ \mu^-$. The predicted branching ratios for these modes, including $\Lambda_b$, are shown in Fig.~\ref{Fig:BR} across five representative $q^2$ bins: $[0.1,\,0.98]~\mathrm{GeV}^2$, $[1.1,\,2.5]~\mathrm{GeV}^2$, $[2.5,\,4]~\mathrm{GeV}^2$, $[4,\,6]~\mathrm{GeV}^2$, and the full low-$q^2$ region $[1,\,6]~\mathrm{GeV}^2$, using the $1\sigma$ ranges of the new Wilson coefficients derived earlier. These regions represent domains where long-distance hadronic uncertainties are moderate, thus allowing relatively clean sensitivity to New Physics. While these additional channels offer useful cross-checks, they are nevertheless subject to similar theoretical uncertainties.  Therefore, to more robustly probe potential NP effects in baryonic decays, we turn to lepton non-universality (LNU) and other physical angular observables, which benefit from reduced sensitivity to hadronic form factors. The ratios $\mathcal{R}_\Lambda$, $\mathcal{R}_\Sigma$, and $\mathcal{R}_\Xi$, shown in Fig.~\ref{Fig:LNU}, offer cleaner probes of NP and allow for a more decisive comparison with the SM expectations. Figures~\ref{Fig:AFBlep}, \ref{Fig:AFBlephad}, and \ref{Fig:AFBhad} showcase the leptonic, leptonic-hadronic, and hadronic forward-backward asymmetries, respectively, for all three channels across the discussed $q^2$ bins. The longitudinal polarization asymmetries for the semileptonic decays of $\Lambda_b$, $\Sigma_b$, and $\Xi_b$ are depicted in Fig.~\ref{Fig:polarization}.  Furthermore, Figs.~\ref{Fig:BR1to6} and \ref{Fig:AFB1to6} provide the predicted $1\sigma$ ranges for various observables of $\Lambda_b \to \Lambda$ (left), $\Sigma_b \to \Sigma$ (middle), and $\Xi_b \to \Xi$ (right) within the low-$q^2$ region $[1,\,6]~\mathrm{GeV}^2$. In particular, Fig.~\ref{Fig:BR1to6} includes the branching ratio (top), LNU parameters (middle), and longitudinal polarization asymmetry (bottom). Meanwhile, Fig.~\ref{Fig:AFB1to6} features the leptonic (top), leptonic-hadronic (middle), and hadronic (bottom) forward-backward asymmetries.




By examining the behavior of physical observables in the SM and under various NP scenarios, we arrive at the following key observations:

\begin{itemize}
    
\item  

In the lowest $q^2$ bin, [0.1–0.98]~GeV$^2$, all NP predictions largely overlap with the SM band, as expected due to the dominance of the photon-pole contribution through $C_7^{\rm eff}$, which suppresses the sensitivity to NP effects in $C_9^{\rm NP}$ and $C_{10}^{\rm NP}$. However, starting from the second bin ([1.1–2.5]~GeV$^2$), distinct deviations among the scenarios begin to emerge. Notably, scenario S-C \big($C_9^{\rm NP}$, $C_{10}^{\rm NP}$\big) exhibits the most significant enhancement in the branching ratios across all three decay modes and in every bin beyond the first. These enhancements reflect constructive interference arising from simultaneous nonzero values of both $C_9^{\rm NP}$ and $C_{10}^{\rm NP}$. Scenario S-D \big($C_9^{\rm NP}$, $C_{10}^{\prime\,\rm NP}$\big) also predicts branching ratios that are consistently above the SM, albeit slightly lower than those of S-C. In contrast, scenarios S-B and S-F, characterized by relations such as $C_9^{\rm NP} = -C_{10}^{\rm NP}$ or inclusion of opposite-sign primed coefficients, lead to branching ratios that are suppressed relative to the SM in the mid- and high-$q^2$ regions. This suppression arises from destructive interference between the NP and SM amplitudes in the differential decay rate. Scenarios S-A (only $C_9^{\rm NP}$) and S-E \big($C_9^{\rm NP} = C_9^{\prime\,\rm NP}$, $C_{10}^{\rm NP} = -C_{10}^{\prime\,\rm NP}$\big) yield results that remain close to the SM or show mild enhancement, but they do not deviate as strongly as S-C or S-D. The $q^2 \in [1,6]~{\rm GeV}^2$ integrated branching ratios preserve the same hierarchy seen in the differential plots, with S-C and S-D showing the strongest enhancement across all decay modes.  The consistent hierarchy among the NP scenarios, S-C predicting the highest branching ratios, followed by S-D, then S-A/S-E, and with S-B and S-F giving the lowest, persists across decay modes and bins. This emphasizes that semileptonic baryonic branching ratios, particularly in the intermediate $q^2$ region, serve as sensitive probes of both the magnitude and chiral structure of possible NP contributions.

\item 

In the SM, the lepton non-universality ratios, defined as the ratio of dimuon to dielectron branching fractions are predicted to be very close to unity across the full $q^2$ range, apart from minor deviations induced by lepton mass effects and associated phase space differences. Any substantial departure from unity therefore constitutes a clean signal of lepton flavor universality violation and may point to the presence of NP that couples differently to muons and electrons. In NP scenarios where the new operators affect only the muon channel, the dielectron decay remains SM-like, while the dimuon decay rate is either enhanced or suppressed, depending on the nature and interference of NP contributions.  Among the benchmark scenarios considered, S-C, involving simultaneous nonzero $C_9^{\rm NP}$ and $C_{10}^{\rm NP}$, exhibits the most significant deviation from unity across all bins and decay modes, particularly in the intermediate $q^2$ regions ([1.1–2.5] and [2.5–4.0]~GeV$^2$), as a result of constructive interference that selectively enhances the dimuon mode. Scenario S-D, with nonzero $C_9^{\rm NP}$ and $C_{10}^{\prime\,\rm NP}$, also yields enhanced LNU ratios, though to a somewhat lesser extent. In contrast, scenarios S-B and S-F, which involve structures such as $C_9^{\rm NP} = -C_{10}^{\rm NP}$ or opposite sign primed contributions, predict LNU ratios that are close to or slightly below unity, due to partial cancellations within the dimuon amplitude. Scenarios S-A (only $C_9^{\rm NP}$) and S-E (with $C_9^{\rm NP} = C_9^{\prime\,\rm NP}$ and $C_{10}^{\rm NP} = -C_{10}^{\prime\,\rm NP}$) show only mild enhancements over the SM expectation. The $q^2\in [1,6]~{\rm GeV}^2$ integrated LNU ratios remain significantly above unity in scenarios like S-C and S-D, highlighting robust sensitivity to lepton flavor nonuniversal NP.
 This behavior can be semi-analytically understood from the dominant terms in the differential decay rate, which for the muon channel scale as $|C_9^{\rm eff} + C_9^{\rm NP}|^2 + |C_{10}^{\rm NP}|^2$, while the electron mode receives only SM contributions. As a result, NP modifies only the numerator of the LNU ratio, selectively amplifying or suppressing the dimuon rate depending on the Wilson coefficient structure. Overall, the results in Fig. \ref{Fig:LNU} highlight the potential of baryonic LNU ratios as powerful observables for probing NP scenarios with lepton flavor dependent couplings.

\item 

The lepton-side asymmetry $A_{\text{FB}}^\ell$, the lepton-hadron correlated asymmetry $A_{\text{FB}}^{\ell B_1}$, and the hadron-side asymmetry $A_{\text{FB}}^{B_1}$ arise from different angular correlations in the fully differential decay distribution and are sensitive to distinct interference structures among the Wilson coefficients.

In the SM, $A_{\text{FB}}^\ell$ exhibits a well-known zero-crossing behavior at low $q^2$, driven by a sign change in the interference term $\text{Re}(C_7^{\rm eff} C_9^{\rm eff*})$, which dominates the numerator. The analytic expression for $A_{\text{FB}}^\ell$ contains interference terms between vector and axial-vector operators, scaling approximately as $\text{Re}(C_9^{\rm eff} C_{10}^{\rm NP*})$ and $\text{Re}(C_7^{\rm eff} C_{10}^{\rm NP*})$ in the presence of NP. The $q^2\in [1,6]~{\rm GeV}^2$ integrated $A_{\text{FB}}^\ell$ values reflect cumulative interference effects, with S-C inducing the most pronounced positive shift relative to the SM.
 Consequently, NP scenarios like S-C and S-D can enhance or shift $A_{\text{FB}}^\ell$ through modified interference patterns, whereas S-B and S-F tend to reduce its magnitude or even flip its sign due to destructive interference.

The lepton-hadron correlated asymmetry $A_{\text{FB}}^{\ell B_1}$, shown in Figure~6, depends on the angular correlation between the lepton and hadron directions. Its semi-analytical expression includes terms such as $\text{Re}(C_9^{\rm eff} C_{10}^{\rm NP*})$ and $\text{Re}(C_7^{\rm eff} C_9^{\rm NP*})$, making it sensitive to both vector and axial-vector NP contributions. Compared to $A_{\text{FB}}^\ell$, this observable is typically smaller in magnitude in the SM and exhibits fewer zero crossings. However, it remains a valuable discriminator, responding distinctly to NP effects. Scenario S-C leads to the most significant enhancement across all decay modes, particularly in the intermediate $q^2$ bins ([1.1–2.5] and [2.5–4.0]~GeV$^2$), while scenarios S-B and S-F predict near-zero or even negative values, owing to cancellations in the interference terms. The $q^2\in [1,6]~{\rm GeV}^2$ averaged $A_{\text{FB}}^{\ell B_1}$ is most sensitive in the $\Xi_b$ mode, where S-C and S-D induce significant deviations from the SM expectation.

The hadron-side asymmetry $A_{\text{FB}}^{B_1}$, defined from the angular distribution of the final state baryon, receives contributions from parity-violating interference terms such as $\text{Re}(C_9^{\rm eff} C_{10}^{\rm NP*})$, $\text{Re}(C_7^{\rm eff} C_{10}^{\rm NP*})$, and helicity-suppressed terms. In the SM, $A_{\text{FB}}^{B_1}$ is generally small but nonzero, with moderate shape variation across $q^2$ bins and decay channels. As with the other asymmetries, S-C produces the largest enhancement, followed by moderate shifts in S-D, while scenarios like S-A and S-E stay close to the SM predictions. Once again, cancellation effects in S-B and S-F reduce or suppress the asymmetry. The $q^2\in [1,6]~{\rm GeV}^2$ integrated values show that S-C consistently enhances the asymmetry across all baryon modes, while S-B and S-F push it closer to zero.

Overall, the combined analysis of $A_{\text{FB}}^\ell$, $A_{\text{FB}}^{\ell B_1}$, and $A_{\text{FB}}^{B_1}$ provides complementary insight into the underlying operator structure. Each asymmetry probes different combinations and phases of the Wilson coefficients, making them powerful observables for disentangling the nature of NP contributions in semileptonic baryonic transitions.

\item 

In the SM, \( F_L \) is determined by the relative strength of longitudinal versus transverse helicity amplitudes and typically lies between 0.5 and 0.75 across most of the $q^2$ range. It tends to be larger at high $q^2$, where the longitudinal contribution dominates due to helicity suppression of transverse modes. The expression for \( F_L \) depends on ratios of squared helicity amplitudes, involving combinations of the Wilson coefficients $C_9^{\rm eff}$, $C_{10}^{\rm NP}$, and $C_7^{\rm eff}$, and thus can be modified in the presence of NP. Among the NP scenarios considered, S-C and S-D lead to the most notable deviations from the SM, enhancing \( F_L \) in the low and intermediate $q^2$ bins. This enhancement is primarily due to the increase in the axial-vector contributions through $C_{10}^{\rm NP}$ and/or $C_{10}^{\prime\,\rm NP}$, which bolster the longitudinal amplitude. On the other hand, S-B and S-F, which introduce cancellation between $C_9^{\rm NP}$ and $C_{10}^{\rm NP}$ or include destructive primed contributions, tend to reduce \( F_L \), especially in the [1.1–2.5] and [2.5–4.0]~GeV$^2$ bins. Scenarios S-A and S-E remain close to the SM predictions throughout. The $q^2\in [1,6]~{\rm GeV}^2$ averaged $F_L$ values retain the qualitative trends of the differential distribution, with clear enhancements in S-C and mild suppression in S-B and S-F.  Overall, the longitudinal polarization fraction provides an important complementary observable, sensitive not just to the magnitude of NP contributions but also to the helicity structure of the underlying operators.
\end{itemize}

\subsection{Sensitivity to Individual Wilson Coefficients}

The previous subsection focused on predictions at six representative NP benchmark points (S-A, S-B, …, S-F) derived from global fits. These benchmarks capture qualitatively distinct NP scenarios, but they correspond to fixed parameter choices. To complement this, we now examine the broader sensitivity of baryonic observables by varying individual Wilson coefficients within their $1\sigma$ ranges. For reference, we reproduce the one-dimensional best-fit (1$\sigma$) values obtained in our earlier analysis~\cite{Mohapatra:2024lmp} in Table~\ref{tab:1Dfits}. 

\begin{table*}[htbp]
\centering
\caption{ Best-fit $[1\sigma]$, pull and p-value($\%$) of different single Wilson coefficient NP scenarios in $b \to s \mu^+ \mu^-$ transitions.}\label{tab:1Dfits}
\begin{tabular}{@{}cccccr@{}}
\toprule[1.2pt] 
Scenario & Coefficient & Best-fit value [$1\sigma$]  & Pull &  p-value ($\%$) \\
\hline
\hline
 \midrule 
1D - I & $C_{10}^{\rm NP}$ & $ 0.26 $ $[\substack{0.36 \\ 0.16}]$& 2.20 & 40.0 \\
1D - II & $C_{9}^{'\rm NP}$ & $ -0.21 $ $[\substack{-0.08 \\-0.33}]$&   1.14 & 38.0\\ 
1D - III & $C_{10}^{'\rm NP}$ & $ 0.02 $ $[\substack{0.11 \\ -0.06}]$   &  0.03 & 13.0\\
1D - IV & $C_9^{\rm NP}=C_{10}^{\rm NP}$ &  $ 0.02 $  $[\substack{0.15 \\ -0.10}]$&   0.16 & 28.0  \\
1D - V & $C_{9}^{'\rm NP}=C_{10}^{'\rm NP}$ &   $ -0.10 $ $[\substack{0.02 \\ -0.24}]$&   0.91 & 30.0  \\
1D - VI & $C_{9}^{'\rm NP}=-C_{10}^{'\rm NP}$ &  $ 0.01 $ $[\substack{0.06 \\ -0.03}]$&   1.02 & 35.0 \\
1D - VII & $C_9^{\rm NP}=-C_9^{'\rm NP}$ & $ -0.30 $ $[\substack{-0.17 \\ -0.43}]$&   2.05 & 39.0 \\
1D - VIII & $C_9^{\rm NP}=-C_{10}^{\rm NP}=-C_{9}^{'\rm NP}=-C_{10}^{'\rm NP}$   & $ -0.12 $ $[\substack{-0.05 \\ -0.18}]$     & 2.03 & 39.0 \\
1D - IX & $C_9^{\rm NP}=-C_{10}^{\rm NP}=C_{9}^{'\rm NP}=-C_{10}^{'\rm NP}$  & $-0.06$ $[\substack{-0.02 \\ -0.09}]$&   2.23  & 40.0 \\
\bottomrule[1.2pt] 
\end{tabular}
\end{table*}
\begin{figure}[htp]
\flushleft
\vspace{-0.8cm}
\hspace*{-1.7cm}\includegraphics[width=18cm,height=6cm]{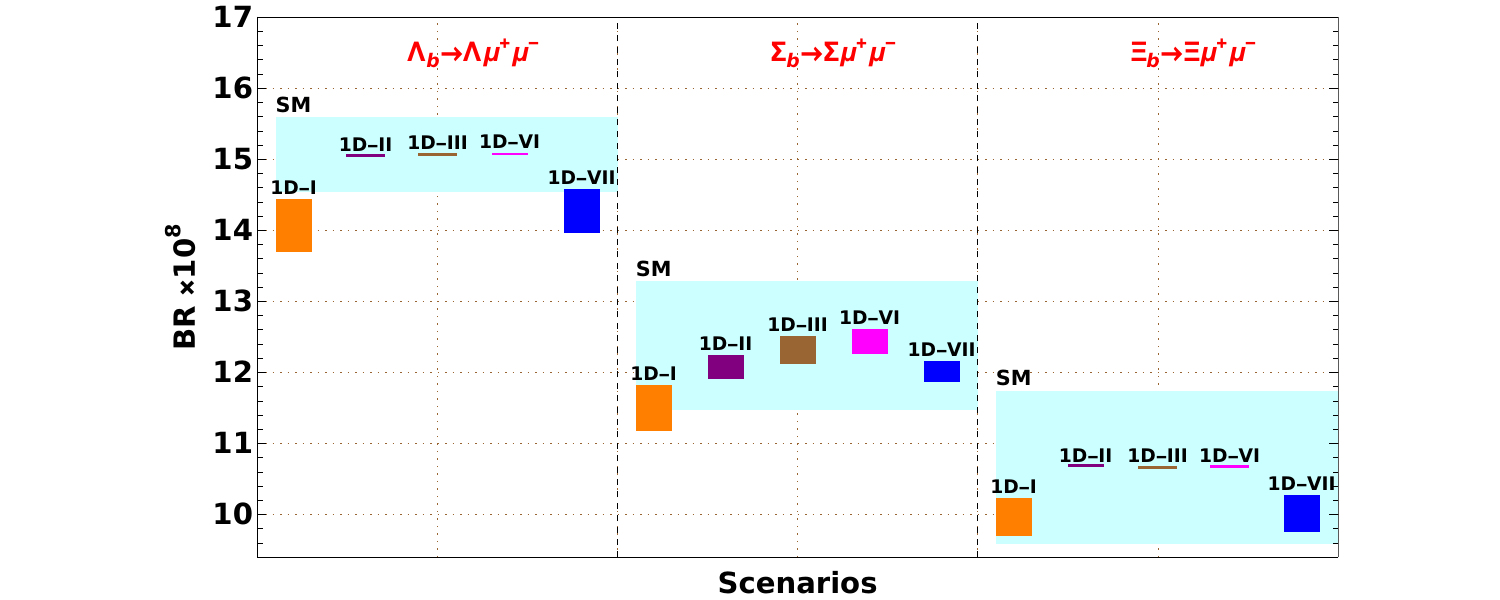}
\caption{Branching ratios of $\Lambda_b \to \Lambda \mu^+ \mu^-$, 
$\Sigma_b \to \Sigma \mu^+ \mu^-$, and $\Xi_b \to \Xi \mu^+ \mu^-$ 
in the low-$q^2 \in [1,6]~{\rm GeV}^2$ region, shown for the 
one-dimensional Wilson coefficient scenarios.}
\label{Fig:add-BR}
\end{figure}
Among all the possible one-dimensional Wilson coefficient scenarios of Table~\ref{tab:1Dfits}, we present the branching ratios of $\Lambda_b \to \Lambda \mu^+ \mu^-$ (left), $\Sigma_b \to \Sigma \mu^+ \mu^-$ (middle), and $\Xi_b \to \Xi \mu^+ \mu^-$ (bottom) in the low $q^2 \in [1,6]~{\rm GeV}^2$ region for five 1D scenarios in Fig.~\ref{Fig:add-BR}. Compared to the SM, the overall deviations are mild in most cases, since the partial rates are largely dominated by hadronic form factors and only indirectly affected by short-distance coefficients. In the $\Lambda_b \to \Lambda \mu^+ \mu^-$ mode, the scenarios $C_{10}^{\rm NP}$ (orange) and $C_9=-C_9'$ (blue) produce clear deviations outside the SM uncertainty band, whereas other scenarios remain consistent with the SM within uncertainties. By contrast, the $\Sigma_b$ and $\Xi_b$ modes stay within uncertainties for all cases considered. The branching ratios therefore serve as valuable consistency checks, but due to sizeable theoretical uncertainties they offer limited diagnostic power for NP. We thus proceed to perform a complementary study of angular and polarization observables, which are theoretically cleaner and potentially more sensitive to short-distance effects.
\begin{figure}[htp]
\flushleft
\vspace{-0.8cm}
\hspace*{-1.7cm}\includegraphics[width=18cm,height=6cm]{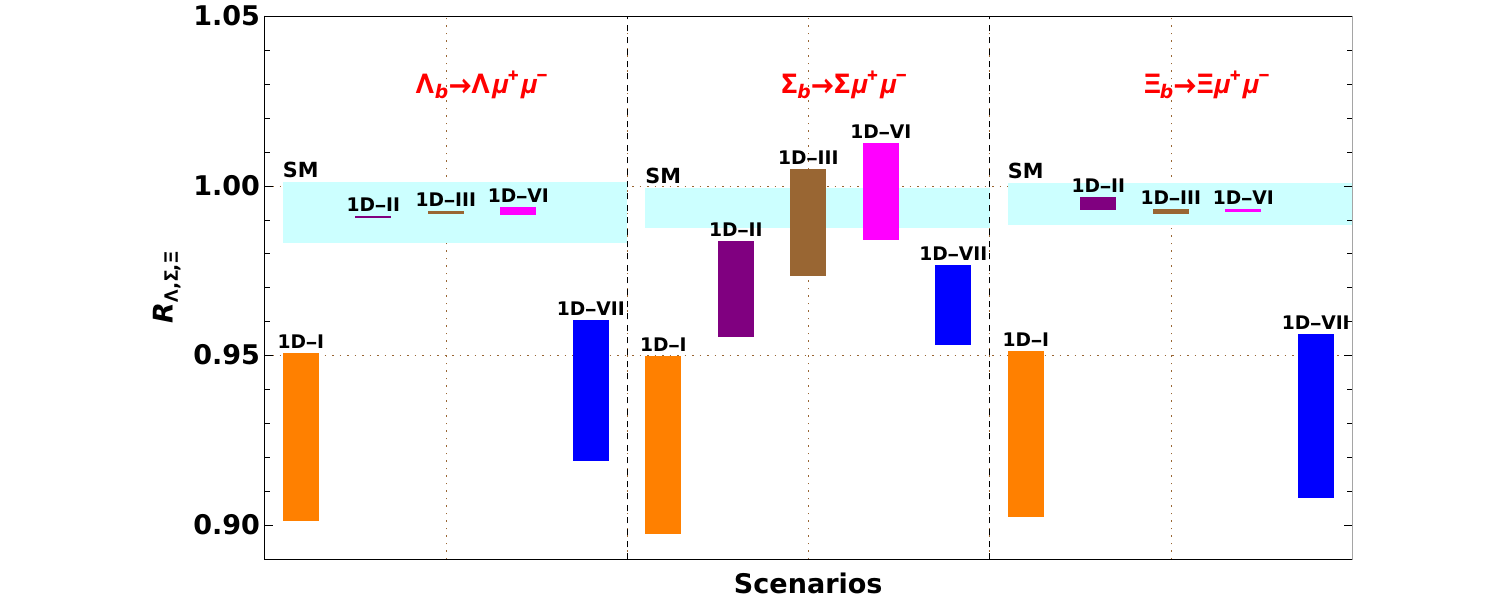}\\
\hspace*{-1.7cm}\includegraphics[width=18cm,height=6cm]{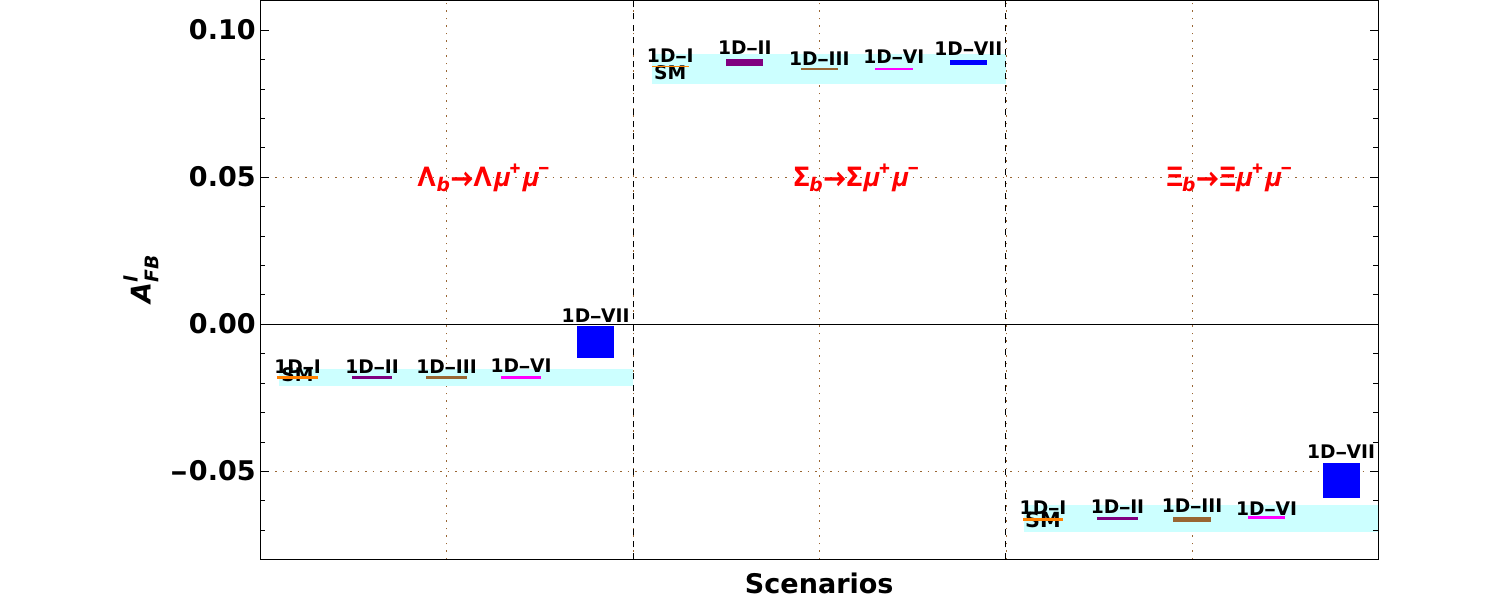}\\
\hspace*{-1.7cm}\includegraphics[width=18cm,height=6cm]{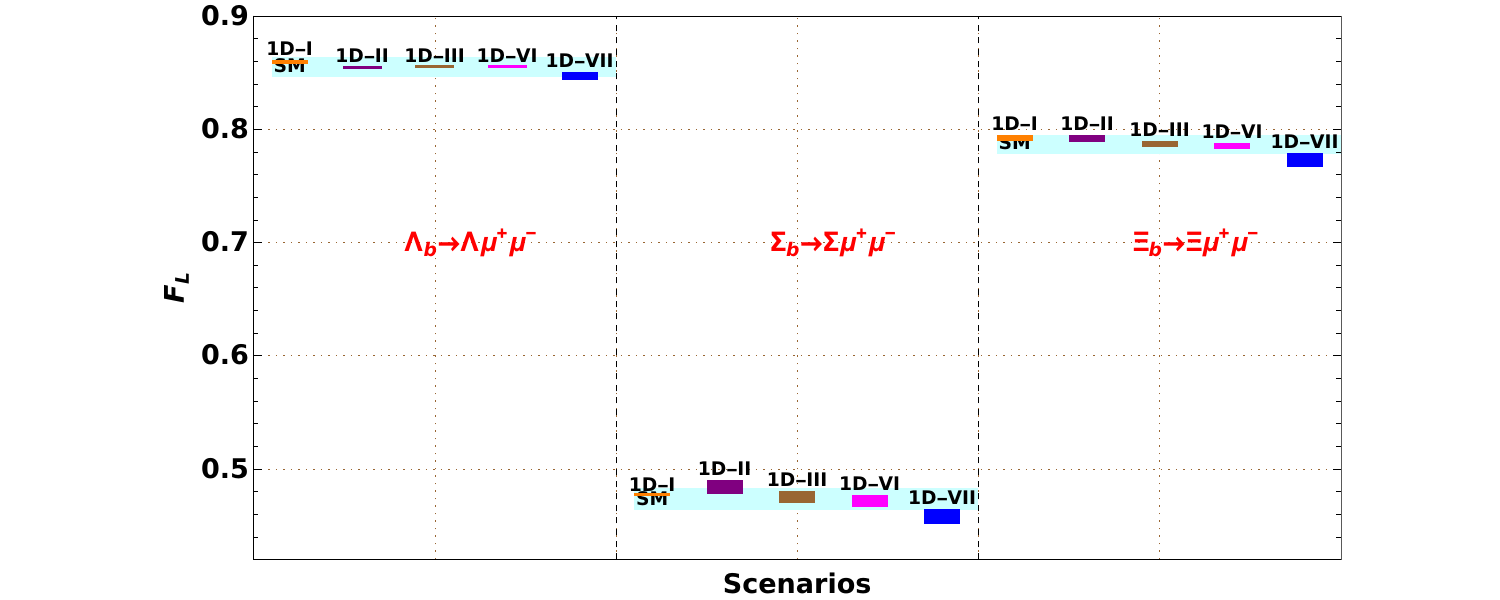}
\caption{Same as Fig.~\ref{Fig:add-BR}, presenting the observables: LNU parameters (top), $\mathcal{A}_{FB}^{\mu}$ (middle), and longitudinal polarization fraction (bottom) in the low $q^2 \in [1,6]~{\rm GeV}^2$ region.}
\label{Fig:add-obs}
\end{figure}

Figure~\ref{Fig:add-obs} displays, for the low-$q^2 \in [1,6]~\mathrm{GeV}^2$ region, the lepton flavor universality ratio $R_\Lambda$ (top), the forward–backward asymmetry $\mathcal{A}_{FB}^{\mu}$ (middle), and the longitudinal polarization fraction $F_L$ (bottom), for the one–dimensional scenarios of Table~\ref{tab:1Dfits}. This figure can be understood from the helicity structure of the transversity amplitudes given in App.~A. Schematically,
\begin{align}
A_i^{L, R} &\propto \Big[(C_{9}^{\rm eff}+C_9\pm C_9') - (C_{10}\pm C_{10}')\Big]\,
      f_i(q^2), 
\end{align}
with $i=\perp,\parallel,0$ labeling the transversity states. The observables then follow as
\begin{align}
\frac{d\Gamma}{dq^2} &\sim \sum_i \left(|A_i^L|^2 + |A_i^R|^2\right), ~
\mathcal{A}_{FB}^\mu &\propto {\rm Re}\left[\sum_i \big(A_i^L A_i^{R*}\big)\right], ~
F_L &\simeq \frac{|A_0^L|^2+|A_0^R|^2}{\sum_i(|A_i^L|^2+|A_i^R|^2)}.
\end{align} 

This structure explains the following numerical behaviour:
\begin{itemize}
  \item \textbf{LNU ratios:}  
  Assuming the electron channel is SM-like, the muon channel receives the NP shifts. The ratios such as $R_{\Lambda, \Sigma_b, \Xi_b}$ cancel much of the hadronic uncertainty and directly reflect the short-distance coefficients. This is why the largest deviations in Fig.~\ref{Fig:add-obs} appear in the LNU panels, especially for 1D-I ($C_{10}^{\rm NP}$) and 1D-VII ($C_9^{\rm NP}=-C_9^{' \rm NP}$) scenarios for $R_{\Lambda_b, \Xi_b}$, while for the $R_{\Sigma_b}$ ratio all the one–dimensional Wilson coefficient scenarios provide significant deviations from the SM. 

  \item \textbf{$\mathcal{A}_{FB}^\mu$:}  
  $\mathcal{A}_{FB}^\mu$ depends on the interference between left and right handed amplitudes. For most 1D scenarios the shifts in $(C_9,C_{10},C_9',C_{10}')$ affect $A_i^L$ and $A_i^R$ in a correlated way, leaving the interference pattern nearly SM-like. The exception is the helicity-flip vector relation $C_9^{\rm NP}=-C_9^{' \rm NP}$, for which
  \[
    C_9^{\rm eff}+C_9^{\rm NP}+ C_9^{' \rm NP}\simeq C_9^{\rm eff},\qquad 
     C_9^{\rm eff}+C_9^{\rm NP}-C_9^{' \rm NP}\simeq C_9^{\rm eff}+ 2\,C_9^{\rm NP} \, .
  \]
  This cancels the left handed vector contribution while enhancing the right handed one, thereby altering the forward–backward asymmetry. This is why only the blue bands deviate visibly in the $\mathcal{A}_{FB}^\mu$ observable. 

  A further important feature is the position of the zero crossing of the forward–backward asymmetry. In the SM this occurs around $q^2 \simeq 4~\mathrm{GeV}^2$, determined approximately by the relation ${\rm Re}[C_9^{\rm eff}(q^2)] \simeq - C_7^{\rm eff}\,\frac{2m_b m_{\Lambda_b}}{q^2}$. Because the zero position depends linearly on the effective $C_9$, it provides a clean probe of NP shifts. As seen in the figure, most one–dimensional scenarios leave the zero crossing nearly unchanged, consistent with their correlated shifts of $C_9$ and $C_{10}$. However, in the helicity-flip case $C_9^{\rm NP}=-C_9^{'\rm NP}$ the cancellation between left- and right-handed amplitudes modifies the effective coefficient controlling the interference, leading to a shift of the zero towards higher $q^2$. This makes the location of the $\mathcal{A}_{FB}^\mu$ zero crossing a complementary observable in distinguishing NP scenarios.

  \item \textbf{$F_L$:}  
  The longitudinal polarization fraction is determined by the relative strength of the longitudinal and transverse amplitudes. Since most NP scenarios shift $C_9$ and $C_{10}$ in a way that rescales left and right handed terms similarly, $F_L$ stays close to its SM value. The exception is again the helicity-flip configuration $C_9^{\rm NP}=-C_9^{'\rm NP}$, which suppresses the left handed vector amplitude while enhancing the right handed one. This distorts the transversity balance and induces a small but visible deviation in $F_L$.  
\end{itemize}

\noindent  In summary, the LNU ratios emerge as the most sensitive probes, since they cleanly expose muon specific short-distance effects, with the strongest impact seen in the $C_{10}^{\rm NP}$ and $C_9^{\rm NP}=-C_9^{\prime \rm NP}$ scenarios. In contrast, $F_L$ and $\mathcal{A}_{FB}^\mu$ show significant deviations only in the helicity-flip case $C_9=-C_9'$, which can be naturally realized in certain $Z'$ models with chiral couplings to quarks and leptons.

\section{Summary and Conclusion}

In this work, we have systematically explored the impact of NP on the semileptonic decays $\Lambda_b \to \Lambda \ell^+ \ell^-$, $\Sigma_b \to \Sigma \ell^+ \ell^-$, and 
$\Xi_b \to \Xi \ell^+ \ell^-$ within a model-independent effective field theory framework. We considered six representative NP scenarios: 
S-A with $C_9^{\text{NP}}$ only; S-B with $C_9^{\text{NP}} = -C_{10}^{\text{NP}}$; 
S-C with both $C_9^{\text{NP}}$ and $C_{10}^{\text{NP}}$ nonzero; 
S-D involving $C_9^{\text{NP}}$ and $C_{10}^{\prime\,\text{NP}}$; 
S-E with $C_9^{\text{NP}} = C_9^{\prime\,\text{NP}}$ and $C_{10}^{\text{NP}} = -C_{10}^{\prime\,\text{NP}}$; and S-F featuring $C_9^{\text{NP}} = -C_9^{\prime\,\text{NP}}$ and $C_{10}^{\text{NP}} = C_{10}^{\prime\,\text{NP}}$. 
We analyzed a wide set of observables, including branching ratios, longitudinal polarization fraction, forward–backward asymmetries ($A_{\text{FB}}^\ell$, $A_{\text{FB}}^{\ell B_1}$, $A_{\text{FB}}^{B_1}$), and lepton non-universality ratios, in both $q^2$-binned and $q^2$-integrated forms.

The branching ratios exhibit strong sensitivity to NP, particularly in S-C and S-D, 
which involve constructive interference among vector and axial-vector operators. 
The lepton non-universality ratios are significantly enhanced in these scenarios as well, serving as clean probes of NP that couples selectively to muons. 
The longitudinal polarization fraction $F_L$ shows noticeable variation in the presence of axial-vector NP, with S-C enhancing and S-B/S-F suppressing it in low and intermediate $q^2$ regions. Angular observables, especially $A_{\text{FB}}^\ell$, exhibit characteristic shifts such as zero-crossing displacements and sign flips under different NP configurations. The combined analysis of $A_{\text{FB}}^{\ell B_1}$ and $A_{\text{FB}}^{B_1}$ further distinguishes between operator structures involving primed coefficients. Overall, our study highlights that semileptonic baryonic decays are sensitive probes of NP, capable of revealing both its magnitude and chiral structure through complementary angular and polarization observables.

Besides these, to further test the sensitivity of baryonic angular observables to different NP structures, we have estimated the branching ratios, forward–backward asymmetry, LNU ratios, and longitudinal polarization asymmetry of these decay modes in the low-$q^2 \in [1,6]~\mathrm{GeV}^2$ region, considering five distinct 
one-dimensional Wilson coefficient scenarios: $C_{10}^{(\prime)\rm NP}$, $C_9^{\prime \rm NP}$, $C_9^{\prime \rm NP}=-C_{10}^{\prime \rm NP}$, and $C_9^{\rm NP}=-C_9^{\prime \rm NP}$. We observe that the $R_{\Lambda}$ and $R_{\Xi}$ ratios exhibit significant deviations in the presence of $C_{10}^{\rm NP}$ and $C_9^{\rm NP}=-C_9^{\prime \rm NP}$, while only a slight deviation appears in the forward–backward asymmetry and the longitudinal polarization fraction for the $C_9^{\rm NP}=-C_9^{\prime \rm NP}$ case. In addition, this scenario also induces a noticeable shift in the zero crossing of the forward–backward asymmetry.

We encourage future experimental efforts at LHCb and upcoming facilities to measure these decay channels with high precision, which will be instrumental in resolving the existing flavor anomalies and probing possible lepton non-universality in the baryonic sector. While some observables, such as branching ratios, remain affected by theoretical uncertainties in hadronic form factors, the angular and polarization observables studied here are relatively clean and offer powerful tools to disentangle different NP scenarios. Current measurements of $\Lambda_b$ decays by LHCb and earlier results from CDF provide a valuable starting point, but remain statistically limited. With the high luminosity upgrades at LHCb and the advent of future olliders, it is expected that many of these observables, particularly those involving angular correlations and lepton non-universality, will become experimentally accessible with higher precision, allowing for stringent tests of flavor dynamics and potential new physics effects.

\section*{Acknowledgment}
AKY extends sincere gratitude to the DST-Inspire Fellowship division, Government of India, for their financial support under ID No. IF210687. MKM acknowledges the financial assistance provided by IoE PDRF from University of Hyderabad.
   
     \appendix

     \section{ The angular coefficients ($\mathcal{K}_i$) for $B_1\to B_2$ decays}\label{A}
The angular coefficients $\mathcal{K}_i$, where $i = 1, 2, 3, 4$, for the decay modes $B_1 \to B_2 \mu^+ \mu^-$ in terms of the transversity amplitudes are given by \cite{Das:2018iap}
     \begin{align}
         \mathcal{K}_{1ss}=&\frac{1}{4}\big(2|A^R_{||0}|^2+|A^R_{||1}|^2+2\,|A^R_{\perp 0}|^2+|A^R_{\perp1}|^2+{R\leftrightarrow L}\big),\nonumber \\
         \mathcal{K}_{1cc}=&\frac{1}{2}\big(|A^R_{||1}|^2+|A^R_{\perp1}|^2+{R\leftrightarrow L}\big),\nonumber \\
         \mathcal{K}_{1c}=&-Re\big(A^R_{\perp1}\,A^{*R}_{||1}-{R\leftrightarrow L}\big),\nonumber \\
         \mathcal{K}_{2ss}=&\frac{\alpha_{B_2}}{2}Re\big(2\,A^R_{\perp0}\,A^{*R}_{||0}+A^R_{\perp1}\,A^{*R}_{||1}+{R\leftrightarrow L}\big),\nonumber \\
\mathcal{K}_{2cc}=&\alpha_{B_2}Re\big(A^R_{\perp1}\,A^{*R}_{||1}+A^L_{\perp1}\,A^{*L}_{||1}\big),\nonumber \\
         \mathcal{K}_{2c}=&-\frac{\alpha_{B_2}}{2}Re\big(|A^R_{\perp1}|^2+|A^R_{||1}|^2-{R\leftrightarrow L}\big),\nonumber \\
         \mathcal{K}_{3sc}=&\frac{\alpha_{B_2}}{\sqrt{2}}Im\big(A^R_{\perp1}\,A^{*R}_{\perp0}-A^R_{||1}\,A^{*R}_{||0}+{R\leftrightarrow L}\big),\nonumber \\
         \mathcal{K}_{3s}=&\frac{\alpha_{B_2}}{\sqrt{2}}Im\big(A^R_{\perp1}\,A^{*R}_{||0}-A^R_{||1}\,A^{*R}_{\perp0}-{R\leftrightarrow L}\big),\nonumber \\
         \mathcal{K}_{4sc}=&\frac{\alpha_{B_2}}{\sqrt{2}}Re\big(A^R_{\perp1}\,A^{*R}_{||0}-A^R_{||1}\,A^{*R}_{\perp0}+{R \leftrightarrow L}\big),\nonumber \\
         \mathcal{K}_{4s}=&\frac{\alpha_{B_2}}{\sqrt{2}}Re\big(A^R_{\perp1}\,A^{*R}_{\perp0}-A^R_{||1}A^{*R}_{||0}-{R\leftrightarrow L}\big),
     \end{align}
     which themselves are expressed in terms of the form factors and Wilson coefficients as follows:
 
     \begin{align}
         A^{L,R}_{\perp1}=&-\sqrt{2}N\Big(f^V_{\perp}\sqrt{2\,s_{-}}\big((C^{eff}_9+C_9^{\rm NP}\mp C_{10}^{\rm NP})+(C^{'\rm NP}_9\mp C^{' \rm NP}_{10})\big)+\frac{2m_b}{q^2}\,f^T_{\perp}(m_{B_1}+m_{B_2})\sqrt{2s_{-}}C^{eff}_7\Big)\,, \nonumber\\
         A^{L,R}_{||1}=&\sqrt{2}N\Big(f^A_{\perp}\sqrt{2\,s_{+}}\big((C^{eff}_9+C_9^{\rm NP}\mp C_{10}^{\rm NP})- (C^{'\rm NP}_9\mp C^{' \rm NP}_{10})\big)+\frac{2m_b}{q^2}\,f^{T5}_{\perp}(m_{B_1}-m_{B_2})\sqrt{2s_{+}}C^{eff}_7\Big)\,,\nonumber\\
         A^{L,R}_{\perp0}=&\sqrt{2}N\Big(f^V_0(m_{B_1}+m_{B_2})\,\sqrt{\frac{s_{-}}{q^2}}\big((C^{eff}_9+C_9^{\rm NP}\mp C_{10}^{\rm NP})+ (C^{'\rm NP}_9\mp C^{' \rm NP}_{10})\big)+\frac{2m_b}{q^2}f^T_0\,\sqrt{q^2\,s_{-}}C^{eff}_7\Big)\,,\nonumber\\
         A^{L,R}_{||0}=&-\sqrt{2}N\Big(f^A_0(m_{B_1}-m_{B_2})\sqrt{\frac{s_{+}}{q^2}}\big((C^{eff}_9+C_9^{\rm NP}\mp C_{10}^{\rm NP})- (C^{'\rm NP}_9\mp C^{' \rm NP}_{10})\big)+\frac{2m_b}{q^2}f^{T5}_0\sqrt{q^2\,s_{+}}C^{eff}_7\Big),
     \end{align} 
     where, $s_{\pm}=(m_{B_1}\pm m_{B_2})^2-q^2$ and the normalization constant $N$ is represented as  
     \begin{align}
         N=\Big[\frac{G^2_F |V_{tb}V^{*}_{ts}|^2\alpha^2_e}{3.2^{11}\,m^3_{B_1}\pi^5}\,q^2\lambda^{1/2}(m^2_{B_1},m^2_{B_2},q^2)\beta_l\Big]^{1/2}\,, \end{align}
         with 
\begin{eqnarray}
\beta_{l} = \sqrt{1 - \frac{4m^2_l}{q^2}}, ~{\rm and}~
\lambda(m^2_{B_1}, m^2_{B_2}, q^2) = m^4_{B_1} + m^4_{B_2} + q^4 - 2(m^2_{B_1}\,m^2_{B_2} + m^2_{B_2}\,q^2 + q^2\,m^2_{B_1}).  \, 
\end{eqnarray}  

Here the SM effective Wilson coefficients $C_7^{\text{eff}}$ and $C_9^{\text{eff}}$ are given as~\cite{Grinstein:2004vb}:
\begin{align}
C_9^{\text{eff}} &= C_9^{\text{SM}} + h(0, q^2) \left( \frac{4}{3} C_1 + C_2 + \frac{11}{2} C_3 - \frac{2}{3} C_4 + 52 C_5 - \frac{32}{3} C_6 \right) \nonumber \\
&\quad - \frac{1}{2} h(m_b, q^2) \left( 7 C_3 + \frac{4}{3} C_4 + 76 C_5 + \frac{64}{3} C_6 \right) \nonumber \\
&\quad + \frac{4}{3} \left( C_3 + \frac{16}{3} C_5 + \frac{16}{9} C_6 \right) + \frac{8 m_c^2}{q^2} \left( \frac{4}{9} C_1 + \frac{1}{3} C_2 + 2 C_3 + 20 C_5 \right) \nonumber \\
&\quad + \frac{\alpha_s}{4\pi} \left[ C_1 \left( B(q^2) + 4 C(q^2) \right) - 3 C_2 \left( 2 B(q^2) - C(q^2) \right) - C_8 F_8^{(9)}(q^2) \right],  \\
C_7^{\text{eff}} &= C_7^{\text{SM}} - \frac{1}{3} \left( C_3 + \frac{4}{3} C_4 + 20 C_5 + \frac{80}{3} C_6 \right) + \frac{\alpha_s}{4\pi} \left[ (C_1 - 6 C_2) A(q^2) - C_8 F_8^{(7)}(q^2) \right],
\end{align}
where $C_7^{\text{SM}}$ and $C_9^{\text{SM}}$ are the SM values evaluated at the $b$-quark mass scale \cite{Ali:1999mm} and the $A(q^2)$, $B(q^2)$, $C(q^2)$, and $F_8^{(7,9)}(q^2)$ loop functions arise in the perturbative QCD corrections to the Wilson coefficients \cite{Beneke:2001at, Seidel:2004jh}.  The $h(m_q, q^2)$ loop function is defined as:
\begin{align}
h(m_q, q^2) &= -\frac{8}{9} \ln\left(\frac{m_q}{m_b}\right) + \frac{8}{27} + \frac{4}{9}x - \frac{2}{9}(2 + x) \sqrt{\abs{1 - x}} \times
\begin{cases}
\ln\left( \frac{\sqrt{1 - x} + 1}{\sqrt{1 - x} - 1} \right) - i\pi, & \text{if } x < 1, \\
2 \arctan \left( \sqrt{\frac{1}{x - 1}} \right), & \text{if } x > 1,
\end{cases}
\label{eq:hq}
\end{align}
where $x \equiv \frac{4 m_c^2}{q^2}$. In the special case of $m_q = 0$, the function simplifies to:
\begin{align}
h(0, q^2) = \frac{8}{27} - \frac{4}{9} \ln\left(\frac{q^2}{m_b^2}\right) + \frac{4}{9} i \pi.
\label{eq:h0}
\end{align}


\section{Form factors of the $B_1\to B_2 l^+ l^-$ modes}\label{B}
\subsubsection{$\Xi_b \to \Xi$  and $\Sigma_b \to \Sigma$  (Light cone QCD sum rule) } 
The expression for the form factors of  $\Xi_b \to \Xi$ \cite{Azizi:2011mw} and $\Sigma_b \to \Sigma$ \cite{Katirci:2012eh} is given by
\begin{eqnarray}
F_i(q^2)=\frac{a}{(1-\frac{q^2}{m_{fit}^2})}+\frac{b}{(1-\frac{q^2}{m_{fit}^2})^2},
\label{parametrization1}
\end{eqnarray}
where, $F= f, g$, $i=1,2,3$. The input values of $a,~b,~m_{fit}$ are presented in Table \ref{Tab:FF1} and \ref{Tab:FF2}. 
\begin{table}[htb]
\caption{Parameters used in the  form factors for $\Xi_{b}\rightarrow \Xi\ell^{+}\ell^{-}$ transition  \cite{Azizi:2011mw}. } \label{Tab:FF1}
\renewcommand{\arraystretch}{1.5}
\addtolength{\arraycolsep}{3pt}
$$
\begin{array}{|c|c|c|c|c|c|c|c|}

\hline  \hline
                & \mbox{a} & \mbox{b}  & m_{fit} & \textcolor{white}{--} & a &b & m_{fit}
                \\
\hline
\hline
 f_1            &   0.166  &  -0.024  &  5.35 &   f_1^{T}   &   0.127  &  -0.129  &  5.10  \\
 f_2            &   0.028  &  -0.048  &  5.31  &  f_2^{T}   &   0.072  &   0.085  &  5.40 \\
 f_3            &  -0.004  &  -0.006   &  5.37 &   f_3^{T}   &  -0.003  &   0.049  &  5.23 \\
 \hline
  g_1            &   0.106   &  0.054  &  5.24  &    g_1^{T}    &   0.288  &  -0.312   &  4.80 \\
 g_2            &  -0.005  &  -0.004   &  5.28  &  g_2^{T}  &   0.036  &   0.119  &  4.70\\
 g_3            &   0.003  &  -0.006   &  4.70  &   g_3^{T}  &   0.024  &  -0.095  &  5.33 \\
\hline \hline
\end{array}
$$
\renewcommand{\arraystretch}{1}
\addtolength{\arraycolsep}{-1.0pt}
\end{table}

\begin{table}[htb]
\caption{Parameters used in the form factors for $\Sigma_{b}\rightarrow \Sigma\ell^{+}\ell^{-}$ decay \cite{Katirci:2012eh}. }
\label{Tab:FF2}
\renewcommand{\arraystretch}{0.9} 
\addtolength{\arraycolsep}{-2pt}  
\small 
\centering
\begin{tabular}{|c|c|c|c|c|c|c|c|}

\hline \hline
                & a & b  & $m_{\text{fit}}$ &  & a & b & $m_{\text{fit}}$
                \\
\hline
 $f_1$            &  $-(0.035\pm0.006)$ &   $0.130\pm0.023$   &  $5.1\pm1.0$ & $f_1^{T}$        &  $1.0\pm0.0$    &   $-(1.0\pm0.0)$   &  $5.4 \pm1.1$ \\
 $f_2$            &  $0.026\pm0.006$  &  $-(0.081\pm0.018)$  &  $5.2 \pm1.0$  &  $f_2^{T}$        &  $-(0.290\pm0.089)$  &   $0.421\pm0.129$   &  $5.4 \pm1.1$  \\
 $f_3$            &   $0.013\pm0.004$ &  $-(0.065\pm0.020)$  &  $5.3 \pm1.1$  & $f_3^{T}$        &  $-(0.240\pm0.071)$  &   $0.412\pm0.122$   &  $5.4 \pm1.1$   \\
 $g_1$            &  $-(0.031\pm0.008)$ &   $0.151\pm0.038$   &  $5.3 \pm1.1$  &  $g_1^{T}$        &  $0.450\pm0.135$   &   $-(0.460\pm0.138)$  &  $5.4  \pm1.1$ \\
 $g_2$            &  $0.015\pm0.005$  &  $-(0.040\pm0.013)$  &  $5.3 \pm1.1$  & $g_2^{T}$        &  $0.031\pm0.009$  &   $0.055\pm0.015$  &  $5.4  \pm1.1$ \\
 $g_3$            &  $0.012\pm0.003$  &  $-(0.047\pm0.012)$  &  $5.4 \pm1.1$  &  $g_3^{T}$        &  $-(0.011\pm0.003)$ &  $-(0.180\pm0.057)$   &  $5.4  \pm1.1$ \\
\hline \hline
\end{tabular}
\renewcommand{\arraystretch}{1} 
\addtolength{\arraycolsep}{1.0pt} 
\end{table}
\subsubsection{$\Lambda_b \to \Lambda$ (Lattice QCD) }
The form factors for $\Lambda_b \to \Lambda$ in lattice QCD \cite{Detmold:2016pkz} are given by 
\begin{equation}
 f(q^2) = \frac{1}{1-q^2/(m_{\rm pole}^f)^2} \big[ a_0^f + a_1^f\:z(q^2) + a_2^f\:[z(q^2)]^2 \big].\label{eq:nominalfitphys}
\end{equation}
The values and uncertainties of the parameters $a_0^f$, $a_1^f$ and $a_2^f$ from the nominal fit are given in Table \ref{tab:HO}.
\begin{table}[ht]
\caption{\label{tab:HO}Parameters used in the form factor for $\Lambda_b \to \Lambda \ell^{+}\ell^{-}$ transition \cite{Detmold:2016pkz}.}
\centering
\begin{tabular}{ccccc}
\hline \hline
Parameter         &  Value  & \hspace{4ex} &  Parameter         &  Value    \\
\hline
$a_0^{f_+}$       & $ 0.4229\pm 0.0274$  & &  $a_2^{g_0}$        & $ 1.1490\pm 1.0327$  \\ 
$a_1^{f_+}$       & $-1.3728\pm 0.3068$     & &  $a_1^{g_\perp}$    & $-1.3607\pm 0.2949$    \\ 
$a_2^{f_+}$       & $ 1.7972\pm 1.1506$  & &  $a_2^{g_\perp}$    & $ 2.4621\pm 1.3711$ \\ 
$a_0^{f_0}$       & $ 0.3604\pm 0.0277$  & &  $a_0^{h_+}$        & $ 0.4753\pm 0.0423$ \\ 
$a_1^{f_0}$       & $-0.9248\pm 0.3453$     & &  $a_1^{h_+}$        & $-0.8840\pm 0.3997$    \\ 
$a_2^{f_0}$       & $ 0.9861\pm 1.1988$  & &  $a_2^{h_+}$        & $-0.8190\pm 1.6760$    \\ 
$a_0^{f_\perp}$   & $ 0.5148\pm 0.0353$  & &  $a_0^{h_\perp}$    & $ 0.3745\pm 0.0313$ \\ 
$a_1^{f_\perp}$   & $-1.4781\pm 0.4030$     & &  $a_1^{h_\perp}$    & $-0.9439\pm 0.2766$    \\ 
$a_2^{f_\perp}$   & $ 1.2496\pm 1.6396$  & &  $a_2^{h_\perp}$    & $ 1.1606\pm 1.0757$ \\ 
$a_0^{g_\perp,g_+}$ & $ 0.3522\pm 0.0205$ & &  $a_0^{\widetilde{h}_{\perp},\widetilde{h}_+}$ & $ 0.3256\pm 0.0248$ \\ 
$a_1^{g_+}$       & $-1.2968\pm 0.2732$     & &  $a_1^{\widetilde{h}_+}$ & $-0.9603\pm 0.2303$ \\ 
$a_2^{g_+}$       & $ 2.7106\pm 1.0665$  & &  $a_2^{\widetilde{h}_+}$ & $ 2.9780\pm 1.0041$ \\ 
$a_0^{g_0}$       & $ 0.4059\pm 0.0267$  & &  $a_1^{\widetilde{h}_\perp}$ & $-0.9634\pm 0.2268$ \\ 
$a_1^{g_0}$       & $-1.1622\pm 0.2929$     & &  $a_2^{\widetilde{h}_\perp}$ & $ 2.4782\pm 0.9549$ \\ 
\hline \hline
\end{tabular}
\end{table}
\bibliographystyle{apsrev4-1}
\bibliography{referance}
\end{document}